\documentclass{aa5}
\usepackage{psfig,latexsym,longtable,lscape}

\def\ros{{\sl ROSAT}}

\def\HI{\hbox{H\,{\sc i}}}

\def\sHI{HI}
\newcommand{\D}{$^\circ$}
\def\p0{\phantom{0}}

\newcommand\approxgt{\mbox{$^{>}\hspace{-0.24cm}_{\sim}$}}

\def\it{\sl}

\begin{document}
   \title{ROSAT X-ray sources in the field of the LMC}
   \subtitle{III.The log N -- log S of background AGN and the LMC gas}
 
   \author{P. Kahabka\inst{1}
           \and K.S. de Boer\inst{1}
           \and C. Br\"uns\inst{2}
          }

   \offprints{P. Kahabka, \email{pkahabka@astro.uni-bonn.de}}
 
   \institute{Sternwarte, Universit\"at Bonn, 
              Auf dem H\"ugel 71, D--53121 Bonn, Germany
              \and Radioastronomisches Institut, Universit\"at Bonn,
              Auf dem H\"ugel 71, D--53121 Bonn, Germany
             }

   \date{Received 27 August 2001/ Accepted 22 April 2002}

\abstract{
We use a sample of 50 background X-ray sources (AGN) and candidate AGN in 
the field of the LMC observed with more than 50 counts in archival \ros\ 
{\sl PSPC} observations to derive the observed $\log N\ -\ \log S$ relation. 
We correct for the inhomogenous \ros\ {\sl PSPC} exposure and the varying 
absorption due to the galactic and the LMC gas (for which we used an \HI\ 
map derived from observations with the {\sl Parkes} radio telescope). We 
compare the observed $\log N\ -\ \log S$ relation with a theoretical 
$\log N\ -\ \log S$ relation of the soft extragalactic X-ray background (SXRB)
which comprises an AGN and a cluster of galaxy contribution. We find that the 
observed $\log N\ -\ \log S$ has a deficiency with respect to the theoretical 
$\log N\ -\ \log S$. There are several factors which can account for such a 
deficiency: (1) incompleteness of the selected AGN and cluster of galaxies 
sample, (2) deviation of the theoretical $\log N\ -\ \log S$ in the LMC field 
from the $\log N\ -\ \log S$ derived from a large sample of AGN in several 
fields in the sky, (3) the existence of gas additional to the \HI\ represented
in the {\sl Parkes} \HI\ map of the LMC field, restricted to the high column 
$\ge 10^{21}\ {\rm cm}^{-2}$ regime. We investigate the likely contribution 
of these effects and find that (1) a fraction (of at most $\sim$$30$\%) of 
the AGN and clusters of galaxies in the LMC field may not have been found in 
our analysis and may contribute to the observed deficiency. The existence of 
extended regions with hot diffuse gas and source crowding makes the detection 
of all AGN and clusters of galaxies very difficult. (2) We cannot exclude a 
deviation of the $\log N\ -\ \log S$ in the field of the LMC from a mean 
theoretical $\log N\ -\ \log S$, especially the cluster of galaxy 
contribution which is of importance in the flux range we are comprising may 
show variations across the sky. (3) If LMC gas in addition to the \HI\ 
represented in the {\sl Parkes} \HI\ map would be responsible for the 
deficiency and if this additional gas is restricted to the high column 
$\ge 10^{21}\ {\rm cm}^{-2}$ regime, and assuming that the metallicity of the 
ISM of the LMC is $-$0.3~dex lower than the metallicity of the galactic ISM,
then a factor of $1.9\pm^{3.3}_{1.6}$ at 90\% confidence of additional gas 
would be required which, if purely molecular, would be equal to a molecular 
mass fraction of $63\pm^{20}_{42}\%$. Such a value would be larger than but 
within the uncertainties consistent with a molecular mass fraction of 
$\sim$$30\%$ derived from CO observations for the high column regime of the 
LMC gas. From this analysis, it follows that some gas additional to the 
measured \HI\ for the high column regime of the LMC gas is likely to be 
required to explain the observed $\log N\ -\ \log S$. But the amount of such 
additional gas is dependent on the completeness of our selected AGN and 
clusters of galaxies sample and on the assumptions made about the description 
of the $\log N\ -\ \log S$ of the SXRB in the field of the LMC. 
}
\titlerunning{The AGN sample and the $ \log N\ -\ \log S$}
\authorrunning{P. Kahabka et al.}
\maketitle
%
\section{Introduction}

Recent deep X-ray studies of selected fields in the sky (with \ros\,
{\sl Chandra} and {\sl XMM}) have shown that the $\log N\ -\ \log S$ 
function of the background sources follows a canonical relation which
does not differ largely across the sky. This finding is supported by the 
fact that at least 70--80\% (and up to 90\%) of the soft extragalactic
X-ray background (SXRB) has been resolved into point-like X-ray sources
(cf. Hasinger et al. 1998, hereafter HBG98; Hasinger et al. 2001; Giacconi 
et al. 2001). Most of these X-ray sources have been identified in optical 
follow-up programs as Active Galactic Nuclei (AGN), i.e. quasars (QSO) and 
Seyfert\,I galaxies (e.g. Schmidt et al. 1998). In addition, at higher 
energies (2--10\,keV), a population of highly absorbed (obscured) AGN is 
required to explain the X-ray background (e.g. Giacconi et al. 2001 and 
references therein). But also clusters of galaxies, groups of galaxies and 
galaxies contribute to the X-ray background. They are in part contained in 
the sample of X-ray background sources detected in deep X-ray surveys (e.g. 
Rosati et al. 1995). The contribution of unobscured and obscured AGN and 
clusters of galaxies to the SXRB as derived from observations has been shown 
to be in agreement with the predictions from the standard model of the cosmic 
X-ray background (e.g. Gilli et al. 1999, hereafter GRS99; Gilli et al. 2001, 
hereafter GSH01).

Numerous \ros\ {\sl PSPC} observations exist of the general field of the Large 
Magellanic Cloud (LMC) which allow a detailed analysis of the source 
statistics in the LMC region. In the present paper we will derive the observed
$\log N\ -\ \log S$ of background X-ray sources in the LMC field and compare 
it with the $\log N\ -\ \log S$ of the SXRB. There exists an observation 
derived $\log N\ -\ \log S$ relation of the SXRB for the Lockman Hole field 
(HBG98). But we will make use of the description of the $\log N\ -\ \log S$ 
of GRS99 and GSH01 which comprises an AGN and a cluster component as we are 
investigating a large field in the sky for which the contribution of clusters 
of galaxies is of importance. We will make use of the description of the 
$\log N\ -\ \log S$ from the ``fast evolution'' model for the cosmic X-ray 
background (see Sec.\,2 for a presentation of the description of the 
$\log N\ -\ \log S$ discussed and used in this paper). One of our goals is to 
derive constraints for the gas between us and the X-ray sources, in particular
for the gas of the LMC. Clearly the statistics of the sources detected is 
influenced by the absorption by intervening gas but it is also influenced by 
the detection capabilities related with the nature of the \ros\ {\sl PSPC}.

The sample of AGN and candidate AGN has been set up in Kahabka (2001, 
hereafter Paper\,II) and is based on the \ros\ {\sl PSPC} catalog of Haberl \&
Pietsch (1999, hereafter HP99). For this AGN sample we will construct in 
Sect.\,3 and 4 the $\log N\ -\ \log S$ which we will correct for the 
variable exposure, the absorption due to the variable LMC $N_{\rm H}$ and 
also for the incompleteness of AGN observed in the sample with a certain 
number of counts in the source circle. The latter incompleteness is due to 
spatial background variations across the merged observations and the 
requirement for the detection of an AGN of an at least $4\sigma$ excess of 
the source counts above the background in the source circle.

We will derive constraints on the LMC gas additional to the \HI\ which 
apparently are required to get agreement between the $\log N\ -\ \log S$ 
derived in this work and the $\log N\ -\ \log S$ of the SXRB inferred in 
fields with very low absorbing columns. We also will allow the cluster 
contribution in the theoretical $\log N\ -\ \log S$ to vary and investigate 
the effect on the constraints for the LMC gas additional to the \HI.
In Sect.\,5 we will derive constraints on the amount of molecular gas under
the assumption that the gas additional to the \HI\ is molecular. In addition 
we will estimate the effect of obscuration of the sky in the LMC field by 
dark clouds.

\section{The theoretical log N -- log S}

From deep X-ray observations in fields at a high galactic latitude it 
has been found that the $\log N\ -\ \log S$ can be well described by a 
powerlaw with different slopes above and below a flux $S_{\rm b}$ 
(cf. Hasinger et al. 1993; HBG98). For fluxes (0.5 -- 2.0~keV) above 
$S_{\rm b} = 2.66\ 10^{-14}\ {\rm erg}\ {\rm cm^{-2}}\ {\rm s^{-1}}$ 
the differential number of sources per flux interval $dN / dS = n(S)$ 
has been determined by HBG98 as 

\begin{equation}
  n(S) = n_{1}\ S_{\small 14}^{-b_{1}}
\end{equation}

\noindent
with $n_{1} = 238.1$ and $b_{1} = 2.72$. Similar values for $n_{1}$ and 
$b_{1}$ are given in Hasinger et al. (1993). The flux $S_{14}$ is given 
in units of $10^{-14}\ {\rm erg}\ {\rm cm^{-2}}\ {\rm s^{-1}}$. From this 
relationship the number $N(>S)$ of sources per square degree and with 
fluxes in excess of a given flux $S$ can be determined from

\begin{equation}
  N(>S) = 138.4\ S_{\small 14}^{-1.72}
\end{equation}

For fluxes (0.5 -- 2.0~keV) below $S_{\rm b}$ $N(>S)$ is given by

 \begin{equation}
  N(>S) = 118.1\ S_{\small 14}^{-0.94}\ -21.36
\end{equation}

This description of the $\log N\ -\ \log S$ has been derived from deep
observations of a small field of 1.4 square degrees in the sky with low 
galactic absorbing columns (the Lockman Hole).

Optical identifications of this 
sample by Schmidt et al. (1998) have shown that a large fraction of the 
X-ray sources is AGN. Other objects in this sample are clusters of galaxies 
and a few foreground stars. The contribution of AGN, clusters of galaxies 
and groups of galaxies to the $\log N\ -\ \log S$ requires a more detailed 
modeling of the $\log N\ -\ \log S$. In addition it has to be taken into 
account that clusters of galaxies and galaxy groups may show variations 
across the sky (cf. Giuricin et al. 2000). For fluxes above 
$10^{-14}\ {\rm erg}\ {\rm cm^{-2}}\ {\rm s^{-1}}$ the contribution of galaxy 
clusters becomes important (cf. GRS99).

Even for fluxes above $\sim$$10^{-13}\ {\rm erg}\ {\rm cm^{-2}}\ 
{\rm sec^{-1}}$ besides AGN clusters of galaxies are already an important 
contribution to the $\log N\ -\ \log S$. As the $\log N\ -\ \log S$ derived 
for the Lockman Hole field does not extend to fluxes in excess of 
$\sim$$10^{-13}\ {\rm erg}\ {\rm sec^{-1}}$ (cf. HBG98) the fraction of 
clusters contained in this field is small ($\sim$$11\%$, cf. Lehmann et al. 
2001).

A more sophisticated model for the description of the SXRB is required
for samples derived from larger fields which extend over at least 
a few square degrees as the cluster contribution to the $\log N\ -\ \log S$ 
is of importance.

Following the description of GRS99 we used in addition to the 
$\log N\ -\ \log S$ the ``flattened'' $\log N\ -\ \log S$
which is the $\log N\ -\ \log S$ modified by a scaling factor $S_{14}^{1.5}$
($S_{14}$ is the flux in units of $10^{-14}\ {\rm erg}\ {\rm cm^{-2}}\
{\rm s^{-1}}$). The Euclidean slope for such a $\log N\ -\ \log S$ would
become horizontal.

From Fig.\,3 of GRS99 we derived an analytical presentation
of the cluster component:

\begin{equation}
  N(>S)_{\rm Cluster} = 
  1.7\times \frac{(2 + \log(S_{\small 14}))^{2.4}}
  {{S_{\small 14}}^{1.5}}
\end{equation}

This presentation of the cluster $\log N\ -\ \log S$ is in the flux range 
$10^{-14}\ {\rm erg}\ {\rm cm}^{-2} {\rm sec}^{-1}$ to $10^{-11}\ {\rm erg}\ 
{\rm cm}^{-2} {\rm sec}^{-1}$ in good agreement with the cluster 
$\log N\ -\ \log S$ derived by Rosati et al. (1995) and De\,Grandi et al. 
(1999).

In order to be consistent with the $\log N\ -\ \log S$ of GRS99 we also 
reproduced the AGN $\log N\ -\ \log S$ from the same Figure (but we will
derive another description for the AGN component to which we will refer later
on). We derive the following analytical presentation (which is a reasonable 
approximation in the flux range $10^{-15}$ to $10^{-11}\ {\rm erg}\ 
{\rm cm}^{-2}\ {\rm sec}^{-1}$).

\begin{equation}
  N(>S)_{\rm AGN} = 
  32\times (\log(120\times S_{\small 14}))^{1.1} \times
  (S_{\small 14})^{-1.7} 
\end{equation}

Recently the ``flattened'' $\log N\ -\ \log S$ extending over the flux 
range $10^{-10}$ to $10^{-15}$ ${\rm erg}\ {\rm cm}^{-2}\ {\rm s}^{-1}$ 
(and below) has been derived from a sample of $\sim$$700$ AGN by Miyaji 
et al. (2000). From such a large AGN sample extending over such a large 
flux range constraints have been derived for the cosmological evolution 
of the soft X-ray selected AGN as a function of the redshift.
There are two main models, pure luminosity evolution (PLE) with redshift
and a luminosity dependent density evolution (LDDE) with redshift. The 
latter model has been found to be consistent with recent observational 
data and is referred to as the ``standard model'' (e.g. GRS99). It 
is ``model A'' of GSH01. This model has been further refined by GSH01 by 
allowing type~2 AGN to evolve faster than type~1 AGN (``fast evolution'' 
and ``model B'' of GSH01). Type~1 and type~2 AGN are in the unification 
scheme of AGN related due to the orientation (with respect to the observer) 
of a molecular torus surrounding the nucleus of the AGN:  Type~1 is not 
obscured while type~2 is obscured.

We derived rough analytical representations of the ``flattened'' 
$\log N\ -\ \log S$ for the standard model (A) and the model with fast
evolution (B) by making a fit to the $\log N\ -\ \log S$ in the flux range 
$10^{-15}$ to $10^{-10}\ {\rm erg}\ {\rm cm}^{-2}\ {\rm s}^{-1}$. But the 
fit is less accurate for the extreme flux ranges around $10^{-10}$ and 
$10^{-15}\ {\rm erg}\ {\rm cm}^{-2}\ {\rm s}^{-1}$ respectively. For model 
(A,B) we used the following analytical representation

\begin{equation}
  S_{14}^{1.5}\ N(>S)_{\rm AGN} = F(S)^{A,B} \times G(S)^{A,B} 
\end{equation}

\noindent
with

\begin{equation}
  F(S)^{A,B} = 
  a_1\times (S_{\small 14})^{a_2}
\end{equation}

\noindent
and

\begin{equation}
  G(S)^{A,B} = C^{A,B}
               \left\{ \begin{array}{r@{\quad:\quad}l}
               e^{-\frac{(\log(S_{\small 14})-a_5)^2}{2(a_3)^2}} & 
               \log(S_{\small 14}) < a_5 \\
                e^{-\frac{(\log(S_{\small 14})-a_5)^2}{2(a_4)^2}} & 
               \log(S_{\small 14}) \ge a_5 \\
               \end{array} \right.
\end{equation}

\noindent
with

\begin{equation}
  C^{A,B} = \frac{10}{\sqrt{2\pi}} 
            \left\{ \begin{array}{r@{\quad:\quad}l}
            a_3/a_4 & \log(S_{\small 14}) < a_5 \\
            1/a_4   & \log(S_{\small 14}) \ge a_5 \\
            \end{array} \right.
\end{equation}

\begin{figure} 
  \centering{
  \vbox{\psfig{figure=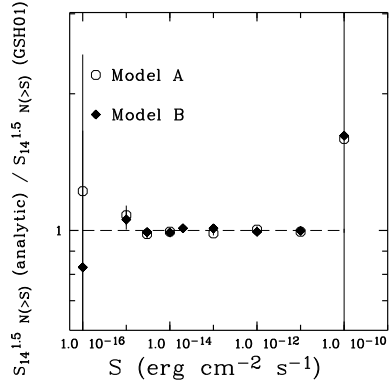,width=5.5cm,angle=0.0,%
  bbllx=0.8cm,bblly=1.5cm,bburx=12.5cm,bbury=12.5cm,clip=}}\par
            }
  \caption[]{Ratio between the analytical presentation and the 
  theoretical model for the ``flattened'' $\log N\ -\ \log S$. The 
  theoretical model (A and B) is taken from Fig.\,3 of GSH01. The analytical 
  presentation is given in Eq.\,6 to 9. The errors given for the points of 
  the model function are the errors used for the least-square fit and are 
  arbitrary. For typical errors of observational data points see Fig.\,3 of 
  GSH01.} 
  \label{ps:modelab}
\end{figure}

The parameters $\rm a_1$, $\rm a_2$, $\rm a_3$, $\rm a_4$ and $\rm a_5$
have been determined from a least-square fit to the curves given in Fig\,3 
of GSH01 for model~A and model~B and are given in 
Table~{\ref{tab:modelabpar}}.

\begin{table} 
     \caption[]{Parameters $a_1$, $a_2$, $a_3$, $a_4$ and $a_5$ for 
     model~A and model~B of the function $S_{\small 14}^{1.5}\ N(>S)$.}
     \begin{flushleft}
     \begin{tabular}{cccccc}
     \hline
     \noalign{\smallskip}
Model & $a_1$  & $a_2$  & $a_3$ & $a_4$ & $a_5$ \\ 
     \noalign{\smallskip}
     \hline
     \noalign{\smallskip}
A     & 109.42 & $-$1.581 & 2.24  & 12.19 & 0.312 \\
B     & 100.2  & $-$1.565 & 2.2   & 12.44 & 0.45  \\
     \noalign{\smallskip}
     \hline
     \end{tabular}
     \end{flushleft}
     \label{tab:modelabpar}
\end{table}

From Eq.\,4 and Eq.\,6 to 9 we can derive the fractional contribution of 
cluster of galaxies to the accumulative number of sources. The fraction of 
clusters of galaxies to the total accumulative source number extends from 
$\sim$47\% to 12\% for the flux range of 
$10^{-12}$ to $10^{-14}\ {\rm erg}\ {\rm cm}^{-2} {\rm s}^{-1}$ respectively
(cf. Tab.\,2).

\begin{table} 
     \caption[]{Cluster to total and cluster to AGN ratio for a given
     limiting flux using Eq.\,4 and Eq.\,6--9, model B.}
     \begin{flushleft}
     \begin{tabular}{cll}
     \hline
     \noalign{\smallskip}
$\log$ flux &$\frac{\rm Cluster}{\rm Cluster + AGN}$  & $\frac{\rm Cluster}{\rm AGN}$ \\
(${\rm erg}\ {\rm cm}^{-2}\ {\rm s}^{-1}$) & \\
     \noalign{\smallskip}
     \hline
     \noalign{\smallskip}
$-12$   & 0.469   & 0.883   \\
$-13$   & 0.269   & 0.367  \\
$-14$   & 0.118   & 0.133  \\
$-15$   & 0.060   & 0.064  \\
     \noalign{\smallskip}
     \hline
     \end{tabular}
     \end{flushleft}
     \label{tab:clusrat}
\end{table}

In Sect.\,4 we will apply this analytical description of the 
``flattened'' $\log N\ -\ \log S$ for the model with fast evolution (B) 
to the observed $\log N\ -\ \log S$ relation. We note that with the 
quality of our data we cannot decide between model~A and B. In 
Fig.~\ref{ps:modelab} we give the ratio of the analytical presentation
of the $\log N\ -\ \log S$ making use of Eq.\,6 to 9 and of the data points 
taken from GSH01.

\section{The completeness correction}

We will discuss effects which determine the incompleteness of the
$\log N\ -\ \log S$ distribution of the AGN sample, the inhomogenous 
{\sl PSPC} exposure across the observed LMC field, the variable absorption 
due to galactic and LMC gas, and the confusion of counts in the source 
circle due to the observation intrinsic background. In Sec.\,4 we will 
correct the observed $\log N\ -\ \log S$ for these incompleteness effects
and compare it with a theoretical $\log N\ -\ \log S$ of the soft 
extragalactic X-ray background.

\subsection{{\sl PSPC} exposure depth}

The merged exposure varies for the LMC field in the range $\sim$(0.6-168)~ksec.
For the AGN given in Paper\,II it follows that an unabsorbed flux of 
$10^{-12}\ {\rm erg}\ {\rm cm^{-2}}\ {\rm s^{-1}}$ in the (0.5 -- 2.0)~keV 
band equals a count rate of $\sim$$0.1\ {\rm s}^{-1}$ in the same energy band. 
Such a value is similar to the count rate of $0.074\ {\rm s}^{-1}$ derived 
from a simulation of an unabsorbed AGN type spectrum.

\begin{figure} 
  \centering{
  \vbox{\psfig{figure=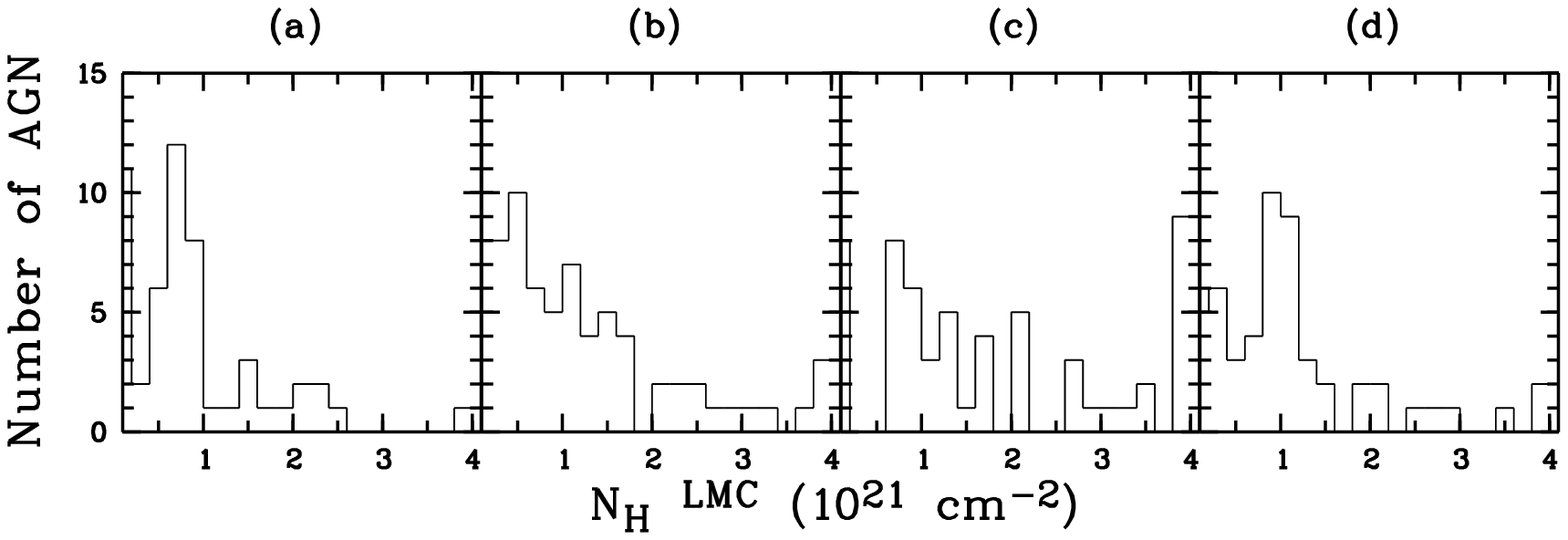,width=8.0cm,angle=0.0,%
  bbllx=0.5cm,bblly=1.8cm,bburx=18.5cm,bbury=8.0cm,clip=}}\par
  \vbox{\psfig{figure=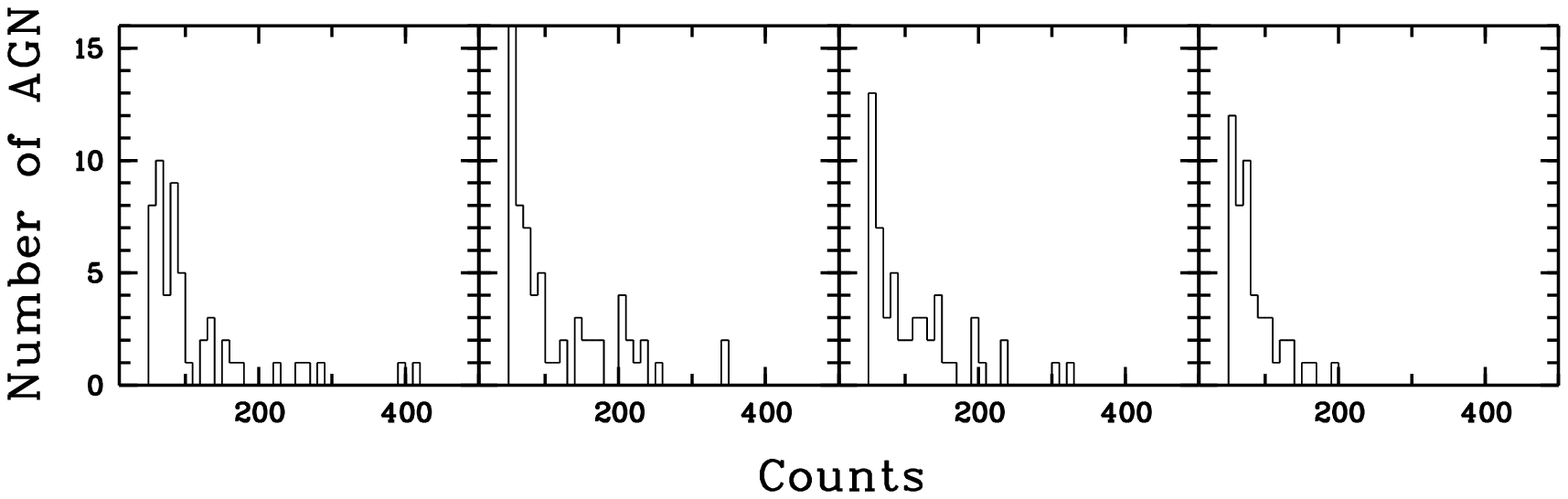,width=8.0cm,angle=0.0,%
  bbllx=0.5cm,bblly=1.8cm,bburx=18.5cm,bbury=7.2cm,clip=}}\par
  \vbox{\psfig{figure=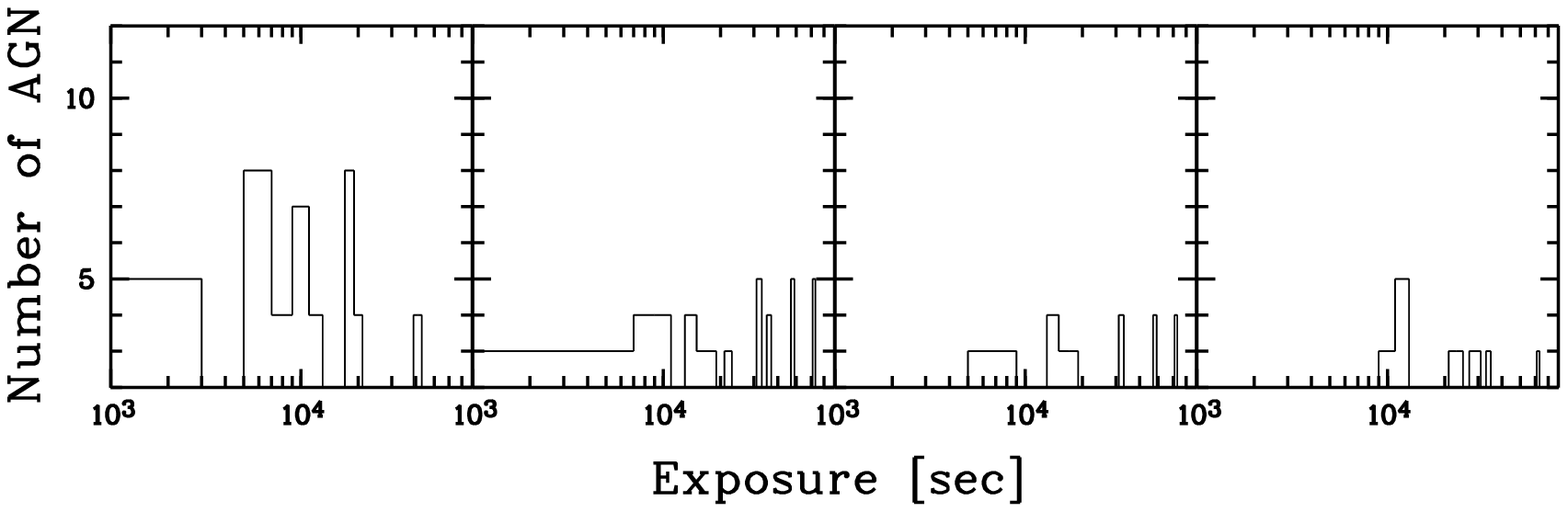,width=8.0cm,angle=0.0,%
  bbllx=0.5cm,bblly=1.8cm,bburx=18.5cm,bbury=7.2cm,clip=}}\par
            }
  \caption[]{Distribution of LMC hydrogen absorbing column density (upper 
  panel), the counts (middle panel) and the exposure (lower panel) 
  of the observed (panel (a)) and simulated (panel (b) to (d)) 
  AGN sample 
  in the investigated LMC field. The samples have been selected for $>$50 
  source counts and have sizes of 50, 71, 57, and 52 sources respectively. 
  The simulated samples are drawn from a sample using an \HI\ map 
  which has $N_{\rm H} = \HI\ $ (b) and which is scaled in the high column 
  $N_{\rm H} > 10^{21}\ {\rm cm}^{-2}$ regime by a factor of two (c)
  respectively. For the simulated sample (d) the cluster scaling factor has 
  been set to 0.2 and an \HI\ map has been used with $N_{\rm H} = \HI\ $.}
  \label{ps:figsim1}
\end{figure}

We have considered for the $\log N\ -\ \log S$ analysis only AGN, for which 
at least 50\,counts have been collected in the source circle making use of 
the merged observations. We assume that such an AGN sample is complete. The 
correctness of this assumption can be confirmed from inspection of the counts 
histogram derived from the observed AGN sample and independently derived 
from a simulated AGN sample (cf. Fig.~\ref{ps:figsim1}).

For a given flux an AGN needs a minimum exposure $E_{min}$ to give the 
required 50 counts. This minimum exposure would be

\begin{equation}
  E_{min} = \frac{675~s}{S_{12}} 
\end{equation}

\noindent
with $S_{12}$ the flux in units of $10^{-12}$ 
${\rm erg}\ {\rm cm}^{-2}\ {\rm s}^{-1}$.
From this equation, it follows that all AGN with fluxes above 
$10^{-12}\ {\rm erg}\ {\rm cm^{-2}}\ {\rm s^{-1}}$ should be contained
in our $\log N\ -\ \log S$ distribution as they require only a 675\,sec 
exposure to be detected. It also follows that faint AGN with a flux of 
$10^{-14}\ {\rm erg}\ {\rm cm^{-2}}\ {\rm s^{-1}}$ would need an exposure
of at least 67\,ksec.

\subsection{Varying \HI\ absorption}

A second effect which contributes to the incompleteness of the observed
AGN sample is the variable absorption due to the galactic and LMC gas. 
Such an absorption further reduces the counts observed from the AGN and 
the fraction of detected AGN for a given flux. In the region of the LMC
both the galactic and the LMC absorbing \HI\ columns show a large variation
(cf. Br\"uns et al. 2001). The largest hydrogen absorbing columns due to 
LMC gas ($>10^{21}\ {\rm cm}^{-2}$) are observed in the eastern cloud 
complex of the LMC. In these regions \ros\ {\sl PSPC} observations with
a deep exposure exist.

We proceeded in the same way as we did when we calculated the incompleteness 
due to the variable exposure across the LMC field. We determined for each
pixel of the merged exposure map the local $N_{\rm H}$ due to the LMC gas
and we added a constant value for the absorption due to the galactic 
foreground gas of $N_{\rm H} = 5\times10^{20}\ {\rm cm}^{-2}$. We 
determined a reduced effective exposure by correcting for the absorption due 
to LMC gas ($N_{\rm H}$ in units of $10^{21}\ {\rm cm}^{-2}$) and due to 
galactic gas (with a column of $5.\ 10^{20}\ {\rm cm}^{-2}$). From simulations
we determined the conversion factor 
$f_{\rm c(band)} = \frac{flux ({\rm erg}\ {\rm cm}^{-2}\ {\rm s}^{-1})}{count 
rate}$. The count rate to flux conversion factor is used to convert the source
count rate to the intrinsic unabsorbed source flux. This flux is used to
construct the $\log N\ -\ \log S$.
For a photon spectrum with a powerlaw index $-\Gamma = 2.0$ and a 
LMC metallicity ($-$0.3~dex) we found for the (0.1 -- 2.4~keV) band

\begin{equation}
  f_{\rm c(0.1-2.4)} = 
  1.0\times10^{-11} + 
  1.64\times10^{-10} \big( (N_{\rm H}+3.9)/20\big)^{1.4}
\end{equation}

\noindent
and for the spectrally hard (0.5 -- 2.0\,keV) energy band

\begin{equation}
  f_{\rm c(0.5-2.0)} = 
  5.7\times10^{-12} + 
  7.7\times10^{-11} \big( (N_{\rm H}+3.9)/20\big)^{1.4}
\end{equation}

Independently we found from simulated unabsorbed \ros\ {\sl PSPC} spectra 
with a powerlaw photon index $-\Gamma = 2.0$, that the intrinsic flux in the
spectrally hard band $f_{c(0.5-2.0)}(N_{\rm H}=0)$ can be determined from the 
intrinsic flux in the spectrally broad band $f_{c(0.1-2.4)}(N_{\rm H}=0)$ as

\begin{equation}
  \frac{f_{c(0.5 - 2.0)}(N_{\rm H}=0)}
       {f_{c(0.1 - 2.4)}(N_{\rm H}=0)}
       = 0.437
\end{equation}

\noindent
(a ratio changing to 0.374, 0.311 and 0.157 for powerlaw slopes $\Gamma$ 
of --2.2, --2.4 and --3.0 respectively). We determined the effective reduced 
exposure $E_{\rm r}$ from the uncorrected exposure $E$ from the equation

\begin{equation}
  E_{\rm r} = {\frac{f_{c}(N_{\rm H}=0)}{f_{c}(N_{\rm H})}} 
  \times E
\end{equation}

Due to absorption by gas in the line of sight to an AGN less counts are 
observed compared to an AGN which is not absorbed. This is equivalent to
a reduced exposure. We determined the LMC $N_{\rm H}$ from an \HI\ map 
measured with the {\sl Parkes} telescope. This map has been aligned using 
reference points. Furthermore we took the variable galactic $N_{\rm H}$ into 
account. We determined the galactic $N_{\rm H}$ from the map of Dickey \& 
Lockman (1991). We determined the flux conversion factor for individual AGN 
by evaluating the {\sl Parkes} 21-cm map for the galactic foreground gas. As 
the count rate to flux conversion factors (cf. Eq.\,11 and 12) have been 
derived for a galactic column of $5\times 10^{20}\ {\rm cm^{-2}}$ we only 
considered the net extra galactic $N_{\rm H}^{\rm gal\ net}\ = 
N_{\rm H}^{\rm gal} - 5\ 10^{20}\ {\rm cm^{-2}}$.

We did not correct the \HI\ values for the LMC gas for self-absorption. But 
the correction factors for self-absorption of the SMC \HI\ have been given 
by Stanimirovic et al. (1999) and from a preliminary analysis of the 
$\log N\ -\ \log S$ of background X-ray sources in the field of the SMC we 
found that consideration of self-absorption has only a small effect on the 
corrected $\log N\ -\ \log S$.

\begin{figure} 
  \centering{
  \vbox{\psfig{figure=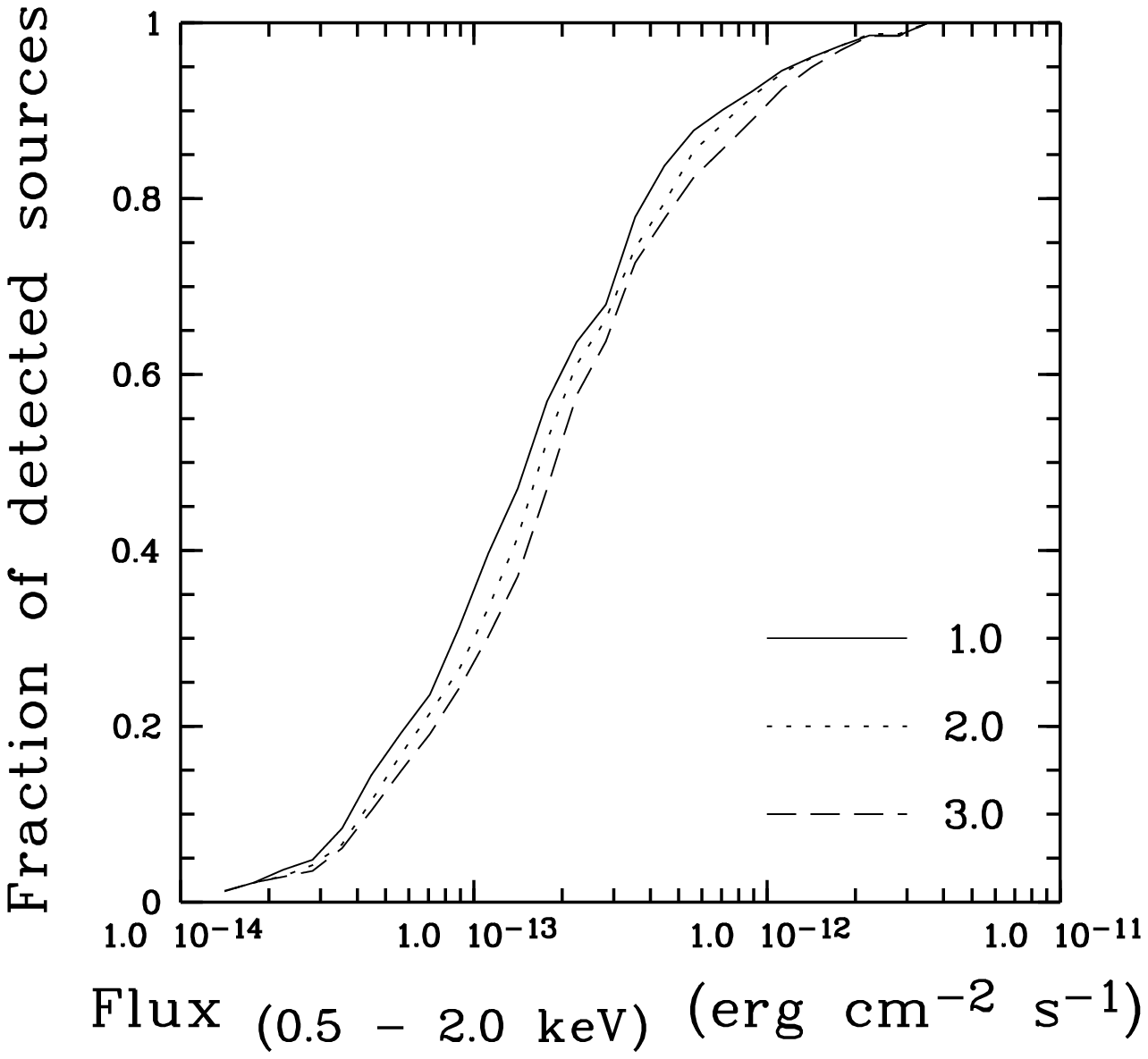,width=4.0cm,angle=0.0,%
  bbllx=2.0cm,bblly=1.5cm,bburx=15.0cm,bbury=13.5cm,clip=}}\par
  \vbox{\psfig{figure=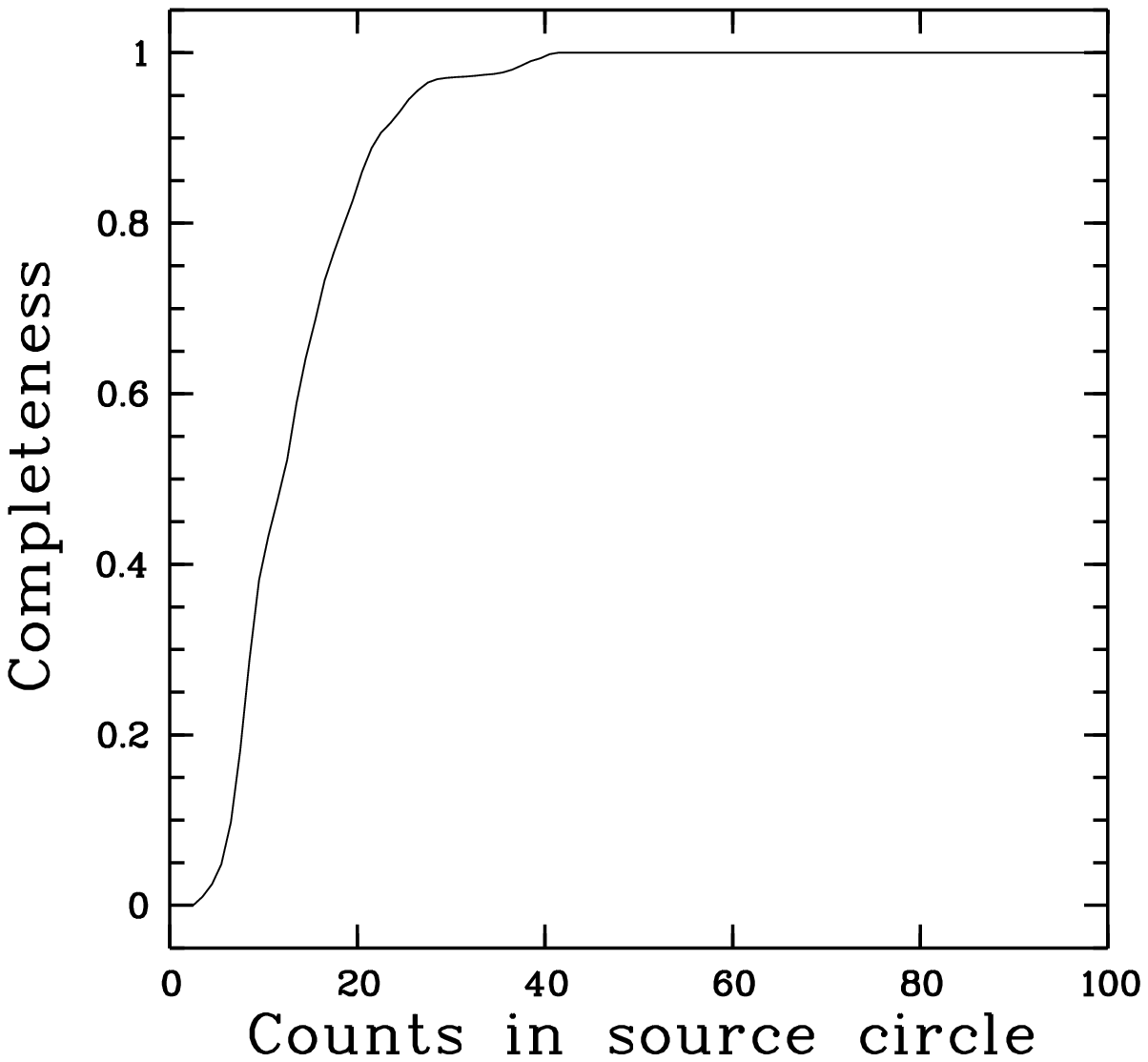,width=4.0cm,angle=0.0,%
  bbllx=2.0cm,bblly=1.5cm,bburx=14.5cm,bbury=13.5cm,clip=}}\par
            }
  \caption[]{Upper panel: The completeness of the AGN sample as a function 
  of the flux and due to the variable merged exposure and the variable 
  absorption due to the LMC gas in the observed LMC field. The fraction 
  of AGN at a given flux and observed with at least 50 counts is given for
  LMC absorbing columns which are scaled by a factor of 1.0, 2.0, and 
  3.0 respectively.
  Lower panel: The completeness of the AGN sample in the LMC field as a 
  function of the counts in the source circle. The completeness has been 
  determined from the 4$\sigma$ excess in counts in the source circle 
  measured from the merged smoothed background image of the observations.}
  \label{ps:lognfrac}
\end{figure}

We also derived from simulations analytical expressions to correct the 
observed {\sl PSPC} count rates into fluxes by taking the absorption due 
to galactic and LMC gas into account. In the upper panel of 
Fig.~\ref{ps:lognfrac} we give the incompleteness (due to the exposure and 
the absorption by the \HI\ gas) of the sample of AGN in the field of the 
LMC which have been observed with more than 50 counts. We give the fraction 
of observed AGN as a function of the flux (0.5 -- 2.0~keV) corrected for 
the absorption due to the intervening galactic and LMC gas.

\subsection{Background confusion} 

A third effect which contributes to the incompleteness of the selected
AGN sample is the confusion of counts in the source circle by counts due 
to the background. We created a merged background image with a binsize
of 1.\arcmin25 from all background images constructed in the spectrally 
hard band and retrieved from the public \ros\ archive. 
Only observations made in the field of the area of the LMC for which the 
$\log N\ -\ \log S$ has been derived have been used. We restricted the 
analysis to the inner 20\arcmin\ of the field of view. We constructed the 
histogram in the 4$\sigma$ excess counts from all observations and we 
determined from this histogram the probability that a source with a given 
number of counts will be detectable in one of the considered observations. 
The completeness of the selected AGN sample as a function of the measured 
counts is given in the lower panel of Fig.~\ref{ps:lognfrac}. It follows 
that for 50~counts the completeness of the AGN sample is 100\% and 
for 20~counts $\sim$85\%.

Part of the observed background is due to extended hot diffuse gas in the
LMC. In these regions it is difficult to detect background AGN due to 
confusion with emission from hot extended gas. Such regions have to be
excluded from the analysis. There is a way to take this effect into account. 
By counting the number of counts from diffuse emission $NCD$ in a cell of the 
size of the source radius one can decide whether a source with $NCS$ counts 
is detectable with more than $\Sigma$ sigma. Using Poissonian statistics one 
finds such a source is detectable only in regions where the number of diffuse 
counts follows the relation 

\begin{equation}
  NCD < (\frac{NCS}{\Sigma})^2
\end{equation}

\begin{figure} 
  \centering{
  \vbox{\psfig{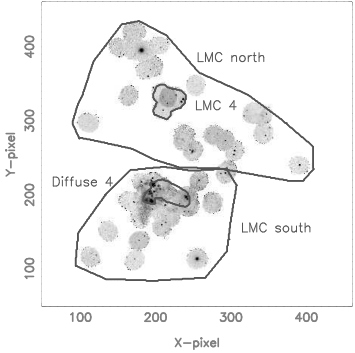}}\par
            }
  \caption[]{The regions {\it LMC north}, {\it LMC south}, and 
  {\it LMC\,4}, which have been used for the $\log N\ -\ \log S$ 
  analysis are marked with dark lines. Also given is the region of 
  extended diffuse emission {\it Diffuse~4} in the 30~Dor complex which 
  has been excluded from the analysis.}
  \label{ps:imalmcreg}
\end{figure}

If we set a significance threshold of $4\sigma$ and a size of the source cell 
of $100\arcsec \times 100\arcsec$ then an AGN with 50 counts can only be 
detected in regions with less than 156 diffuse counts in the source radius. 
We thus excluded from the $\log N\ -\ \log S$ analysis regions of the 
LMC where more than 156 diffuse counts are detected in the spectrally hard 
(0.5 -- 2.0~keV) band and considered the merged exposure of the inner 
20\arcmin\ of the detector. This region is located around the 30\,Dor complex
(cf. Fig.~\ref{ps:imalmcreg}).

\section{The fit of the observed log N -- log S}

For a threshold of 50 counts for the AGN sample given in Paper\,II, 53 
AGN are found in the field of the LMC. If we exclude the region of extended 
diffuse emission Diffuse~4 then we find 50 AGN in the same field. We will
apply the $\log N\ -\ \log S$ analysis in the LMC field to this sample of 
50 AGN. But we will choose a lower threshold of 30 counts for the 
$\log N\ -\ \log S$ analysis applied to the AGN sample derived for the 
field of the Supergiant Shell LMC\,4. We constructed, for the sample of 50 
AGN, the observed $\log N\ -\ \log S$ which we corrected for the incompleteness
due to the varying exposure depth, the varying absorption due to the galactic 
and LMC \HI\ gas, and for background confusion. We excluded the extended 
region of hot diffuse gas in the 30~Dor complex. We derived the 
$\log N\ -\ \log S$ function from the count rates measured in the spectrally 
hard and broad band. But in the $\log N\ -\ \log S$ function the flux has 
always been corrected to that of the spectrally hard band in order to allow 
a direct comparison.

\begin{figure}
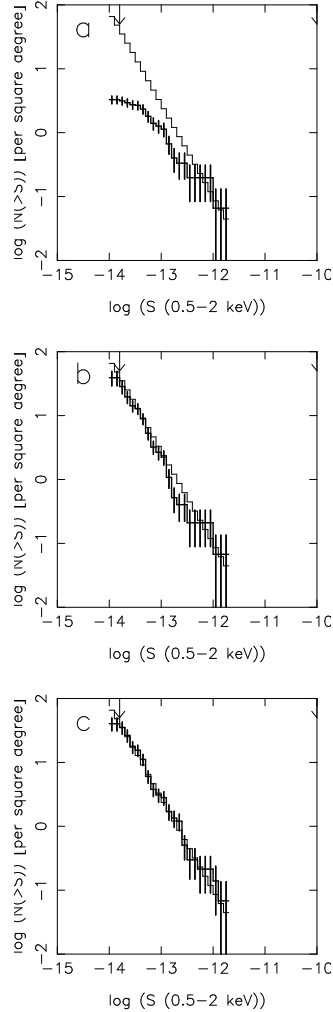
 
  \centering{
  \vbox{\psfig{figure=MS1834f5a.eps,width=4.2cm,angle=-90.0,%
  bbllx=3.5cm,bblly=4.0cm,bburx=19.5cm,bbury=18.6cm,clip=}}\par
  \vbox{\psfig{figure=MS1834f5b.eps,width=4.2cm,angle=-90.0,%
  bbllx=3.5cm,bblly=4.0cm,bburx=19.5cm,bbury=18.6cm,clip=}}\par
  \vbox{\psfig{figure=MS1834f5c.eps,width=4.2cm,angle=-90.0,%
  bbllx=3.5cm,bblly=4.0cm,bburx=19.5cm,bbury=18.6cm,clip=}}\par
            }
  \caption[]{$\log N\ -\ \log S$ of the classified AGN sample in the 
  LMC field derived from the observed count rates in the spectrally hard 
  (0.5 -- 2.0~keV) band for off-axis angles $<$19.\arcmin5. The chosen 
  sample of AGN is complete for $\ge 50$ counts. (a) The 
  $\log N\ -\ \log S$ has not been corrected for incompleteness. (b) 
  The $\log N\ -\ \log S$ has been corrected for incompleteness 
  due to the exposure and the absorbing galactic and LMC gas assuming 
  $N_{\rm H} = N_{\rm HI}$. (c) The $\log N\ -\ \log S$ 
  corrected for incompleteness due to the variable exposure across the LMC 
  field and corrected for the incompleteness due to absorption by galactic 
  and LMC gas (the \HI) across the field but with the LMC 
  $N_{\rm H}$ scaled by a factor 3.2. The thin solid histogram gives the 
  theoretical $\log N\ -\ \log S$ (model B of GSH01) and the arrow marks 
  the lower flux threshold used in the least-square fit.}
  \label{ps:logncorr}
\end{figure}

In Fig.~\ref{ps:logncorr}a we show for the AGN 
sample the uncorrected $\log N\ -\ \log S$ relation in the spectrally 
hard band. In the further investigation we always will show the 
$\log N\ -\ \log S$ in the spectrally hard band.
We now can assume that the discrepancy between the observed 
$\log N\ -\ \log S$ and the standard $\log N\ -\ \log S$
shows us the incompleteness due to exposure depth and \HI\ absorption
variations.

We assume that we have considered all AGN at a given flux which have been 
detected with at least 50 counts. From inspection of
Fig.~\ref{ps:lognfrac} it follows that the completeness of our AGN sample
is expected to be better than 90\% for fluxes above $10^{-12}\ {\rm erg}\
{\rm cm}^{-2}\ {\rm sec}^{-1}$. For fluxes of $10^{-12}$ to $10^{-13}\ 
{\rm erg}\ {\rm cm}^{-2}\ {\rm sec}^{-1}$ the incompleteness should be
not larger than a factor of three while at the low flux end of
$10^{-14}\ {\rm erg}\ {\rm cm}^{-2}\ {\rm sec}^{-1}$ the observed AGN
sample is complete to less than 10\%.

The $\log N\ -\ \log S$ corrected for the variable exposure and the \HI\ 
absorption is given in Fig.~\ref{ps:logncorr}b.
Surprisingly it is found that this $\log N\ -\ \log S$ corrected for 
the variable exposure in the LMC field and the variable intervening galactic
and LMC gas is below the  $\log N\ -\ \log S$ of the SXRB (e.g. using
the description from model~B of GSH01). There could be several reasons for 
this discrepancy.

a) There is the possibility of the existence of gas additional to the
(galactic and LMC) \HI\ derived from a {\sl Parkes} 21-cm survey of the 
field of the LMC. Such additional gas which is not reproduced in the \HI\
data may be in the form of atomic or molecular gas. In Sect.\,5.2 of Kahabka 
et al. (2001, hereafter Paper\,I) we have investigated likely systematic 
effects to derive \HI\ columns with the {\sl Parkes} beam compared with a 
much narrower beam (the {\sl ATCA} beam). We investigated the gas of the 
Small Magellanic Cloud (SMC) and we found that for \HI\ columns below 
$\sim$$3.5\ 10^{21}\ {\rm cm^{-2}}$ the \HI\ columns may be systematically
underestimated by $\sim$$(5-15)\%$. The additional gas aspect will be
explored in Sect.\,4.1.

b) There is also the possibility that the $\log N\ -\ \log S$ in 
the field of the LMC deviates from the $\log N\ -\ \log S$ of the SXRB.
The $\log N\ -\ \log S$ is in our description from Sect.\,2 composed
of two components, an AGN and a cluster component. We here assume that the
description of the AGN component is according to model B of GSH01. The
cluster component of the $\log N\ -\ \log S$ which we derive from GRS99
will in Sect.\,4.2 be allowed to vary. We introduce a scaling factor for 
the cluster component and we explore the constraints on the LMC gas and 
the $\log N\ -\ \log S$.

\subsection{Constraining intervening LMC gas columns}

Assuming there exists additional gas in the line of sight towards the AGN 
which is not contained in the used \HI\ column density map and which has 
not been accounted for in constructing the $\log N\ -\ \log S$, we can 
rescale the \HI\ value in the direction towards the AGN, thus constrain 
such additional gas from a least-square fit (of the observed 
$\log N\ -\ \log S$ compared to a theoretical $\log N\ -\ \log S$).
We simplified the numerical effort as we applied it only to \HI\ columns 
in excess of $10^{21}\ {\rm cm}^{-2}$. We also used a threshold of 50 
source counts.

Other systematic effects which we will take into account in this section are 
due to differences in the abundance models and in the AGN spectral models. 
There exist different abundance models for the interstellar medium (ISM). Also
the details of the metal abundance in LMC gas are not accurately known. The 
average metallicity of LMC gas is, however, known to be approximately $-0.3$ 
dex below the galactic metallicity (de Boer 1991; Russell \& Dopita 1992). 
To calculate the absorption of X-rays the total metal content is needed, 
irrespective of the split between gas-phase and dust-depleted metals. 
Originally, models were based on the metallicity scale of Morrison \& McCammon 
(1983, MM). Recently Wilms, Allen \& McCray (2000, WAM) have updated the 
photoabsorption cross sections in the X-ray regime. In addition they have 
presented a set of abundances for the ISM taking the gas and the dust phase 
into account. We will make use of the abundance models of MM and WAM and we 
will assume for the LMC gas a mean metallicity of $-$0.3~dex.

\begin{table} 
     \caption[]{Count rate to flux conversion factors for models with
     different sets of abundances. We give for each model the abundances 
     $A_{\rm Z}$ for element $Z$ in the nomenclature $12 + \log A_{\rm Z}$. 
     We give the count rate to flux conversion factors $f_{\rm c}$ 
     for the spectrally hard and the spectrally broad \ros\ {\sl PSPC} band. 
     The conversion factor $f_{\rm c}$ is determined by the equation
 $f_{\rm c} = 10^{-11} \times (f_1 + f_2 \times (N_{\rm H} + 3.9)/20)^{1.4})$.
     The coefficients $f_1$ and $f_2$ are given for each model and
     each spectral band.}
     \begin{flushleft}
     \begin{tabular}{lcc|cc}
     \hline
     \noalign{\smallskip}
 &$-\Gamma$ &$A_{\rm Z}$ & $f_{1}$ & $f_{2}$                         \\
     \noalign{\smallskip}
     \hline
     \noalign{\smallskip}
\multicolumn{3}{c}{Model}&\multicolumn{2}{c}{broad (0.1 -- 2.4 keV)} \\
     \noalign{\smallskip}
     \hline
     \noalign{\smallskip}
Ba & 2.0& MM  & 1.00 & 16.4 \\
Bb & 2.0& WAM & 1.45 & 12.8 \\
Bc & 2.2& MM  & 1.00 & 20.5 \\
Bd & 2.2& WAM & 1.50 & 16.2 \\
Be & 2.4& MM  & 1.00 & 26.0 \\
Bf & 2.4& WAM & 1.70 & 20.5 \\
     \noalign{\smallskip}
     \hline
     \noalign{\smallskip}
\multicolumn{3}{c}{Model}&\multicolumn{2}{c}{hard (0.5 -- 2.0~keV)} \\
     \noalign{\smallskip}
     \hline
     \noalign{\smallskip}
Ha & 2.0& MM  & 0.57 & 7.7 \\
Hb & 2.0& WAM & 0.77 & 6.0 \\
Hc & 2.2& MM  & 0.52 & 8.2 \\
Hd & 2.2& WAM & 0.70 & 6.5 \\
He & 2.4& MM  & 0.46 & 8.7 \\
Hf & 2.4& WAM  & 0.70 & 6.8 \\
     \noalign{\smallskip}
     \hline
     \end{tabular}
     \end{flushleft}
In the notation for the model, (B) and (H) refer to the broad and hard 
spectral band respectively, (a,c,e) and (b,d,f) for the abundance 
model of MM and WAM respectively (with MM, Morrison \& McCammon, 1983, and 
$-$0.3~dex for LMC abundances (C=8.35, N=7.39, O=8.57, Ne=7.84, Na=6.02, 
Mg=7.30, Al=6.19, Si=7.27, S=6.98); WAM, Wilms, Allen, \& McCray, 2000, 
and $-$0.3~dex for LMC abundances (C=8.08, N=7.58, O=8.39, Ne=7.64, 
Na=5.86, Mg=7.10, Al=6.03, Si=6.97, S=6.79). In addition a,c,e (and b,d,f)
refer to $-\Gamma$ values of $2.0, 2.2$, and $2.4$ respectively.
     \label{tab:defmod}
\end{table}

The count rate to flux conversion factor $f_{\rm c}$ derived from simulations
in Sec\,3.2 depends on the chosen set of abundances. We here derived 
$f_{\rm c}$ for the set of abundances of Morrison \& McCammon (1983) and WAM, 
cf. Tab.\,3.

The flux distribution of AGN can, in most cases, be described by a powerlaw 
model in the energy range of the \ros\ {\sl PSPC}. The photon powerlaw index 
$\Gamma$ covers values which lie in a narrow range (cf. Paper\,I). The mean 
value of $-\Gamma$ is $\sim$2.2 with a 1$\sigma$ deviation of $\sim$0.2. In 
several investigations it has been found that AGN type spectra cover 
a narrow range in $\Gamma$ with the ``extreme'' range of $-\Gamma$ values 
extending from $2.0$ to $3.0$ respectively (cf. Brinkmann et al. 2000, and 
discussion in Paper\,I). The count rate to flux conversion factor
$f_{\rm c}$ has in addition been derived for spectral models with powerlaw 
indices in the range $-\Gamma = (2.0 - 3.0)$, cf. Tab.\,3.

We restricted the further $\log N\ -\ \log S$ analysis to the abundance 
models and the models for the AGN spectral flux distribution given in 
Tab.\,3. In addition, the flux conversion factors have been calculated for 
two different spectral bands, the hard (0.5 -- 2.0~keV) and the broad
(0.1 -- 2.4~keV) band. We applied a least-square fit to the observed 
$\log N\ -\ \log S$ which has been corrected for incompleteness and 
assuming different sets of conversion factors.

The gas fractions additional to the \HI\ constrained from a least-square 
fit to the $\log N\ -\ \log S$ of background X-ray sources in different 
areas of the LMC field are given in Table~{\ref{tab:gasfrac}}. In Column 
(1) of the table the field designation is given, in Column (2) the area of 
the field, in Column (3) the spectral model used, in Column (4) the counts 
threshold, in Column (5) the used flux range, in Column (6) and (7) the
number of AGN considered (in the whole and high column regime of the LMC gas),
in Column (8) the gas fraction additional to the \HI\ derived from the 
least-square fit, and in Column (9) the reduced chi-squared of the fit.
The result for additional gas averaged over the different models and derived 
for the spectrally hard band is in the range 
$N_{\rm H}^{\rm LMC} = 1.9\pm^{3.3}_{1.6}\ \HI$. The amount of gas in excess 
of the observed \HI\ derived for the spectrally broad band is 
$N_{\rm H}^{\rm LMC} = 1.5\pm^{3.7}_{1.1}\ \HI$.The adjusted 
$\log N\ -\ \log S$ derived from the classified AGN sample making use 
of the spectrally hard flux is given in the lower panel of 
Fig.~\ref{ps:logncorr}. We obtained a generally good match with the standard 
$\log N\ -\ \log S$ relation.

\begin{table*}[htbp]
     \caption[]{Fraction $n$ of gas additional to the \HI\ in the high column 
     ($>10^{21}\ {\rm cm^{-2}}$) regime derived from a $\log N\ -\ \log S$ 
     analysis for different fields in the LMC (90\% confidence). The designation 
     for the field (cf. Table~{\ref{tab:defmod}}), the area of the field, the 
     spectral model (cf. Table~{\ref{tab:defmod}}), the counts threshold, the 
     flux range, the number of AGN used in the $\log N\ -\ \log S$ analysis, and
     the reduced $\chi^2$ and the degrees of freedom (DOF) are given.}
     \begin{flushleft}
     \begin{tabular}{ccccccccc}
     \hline
     \noalign{\smallskip}
Field & Area & Model & Counts & Flux &\multicolumn{2}{c}{Number}& gas & $\chi^{2}_{\rm red}$/DOF  \\
        &   &    & threshold & range &\multicolumn{2}{c}{AGN}   & fraction & \\
        &($\Box$\D) & & & (log (${\rm erg}\ {\rm cm^{-2}}\ {\rm s}^{-1}$)) &(a)&(b)& n & \\
     \noalign{\smallskip}
     \hline
     \noalign{\smallskip}
LMC field &15.2&Ha&50&(-13.8,-11.0)&50&16&1.8$\pm^{3.4}_{1.6}$ &0.42/20\\
LMC field &15.2&Hb&50&(-13.8,-11.0)&50&16&2.4$\pm^{3.0}_{1.8}$ &0.44/20\\
LMC field &15.2&Hc&50&(-13.8,-11.0)&50&16&1.6$\pm^{3.4}_{1.4}$ &0.42/20\\
LMC field &15.2&Hd&50&(-13.8,-11.0)&50&16&2.2$\pm^{3.2}_{1.8}$ &0.37/20\\
LMC field &15.2&He&50&(-13.8,-11.0)&50&16&1.4$\pm^{3.2}_{1.4}$ &0.47/20\\
LMC field &15.2&Hf&50&(-13.8,-11.0)&50&16&2.0$\pm^{3.4}_{1.6}$ &0.44/20\\
LMC field &15.2&Ba&50&(-13.8,-11.0)&50&16&1.4$\pm^{3.8}_{1.0}$ &0.50/21\\
LMC field &15.2&Bb&50&(-13.8,-11.0)&50&16&1.8$\pm^{3.6}_{1.2}$ &0.55/21\\
LMC field &15.2&Bc&50&(-13.8,-11.0)&50&16&1.2$\pm^{3.8}_{1.0}$ &0.45/21\\
LMC field &15.2&Bd&50&(-13.8,-11.0)&50&16&1.8$\pm^{3.6}_{1.2}$ &0.61/21\\
LMC field &15.2&Be&50&(-13.8,-11.0)&50&16&1.4$\pm^{3.6}_{1.2}$ &0.50/20\\
LMC field &15.2&Bf&50&(-13.8,-11.0)&50&16&1.6$\pm^{3.8}_{1.0}$ &0.54/21\\
LMC field north&8.69&Ha&50&(-13.7,-11.0) &34&6& 2.6$\pm^{2.8}_{2.4}$ &0.90/18\\
LMC field north&8.69&Hb&50&(-13.7,-11.0) &34&6& 3.4$\pm^{2.0}_{2.6}$ &0.92/18\\
LMC field north&8.69&Hc&50&(-13.7,-11.0) &34&6& 2.4$\pm^{3.0}_{2.4}$ &0.89/18\\
LMC field north&8.69&Hd&50&(-13.7,-11.0) &34&6& 3.2$\pm^{2.2}_{2.8}$ &0.91/18\\
LMC field north&8.69&He&50&(-13.7,-11.0) &34&6& 2.2$\pm^{3.2}_{2.6}$ &0.85/18\\
LMC field north&8.69&Hf&50&(-13.7,-11.0) &34&6& 3.0$\pm^{2.4}_{2.4}$ &0.92/18\\
LMC field north&8.69&Ba&50&(-13.7,-11.0) &34&6& 2.2$\pm^{3.2}_{2.6}$ &0.44/21\\
LMC field north&8.69&Bb&50&(-13.7,-11.0) &34&6& 2.0$\pm^{3.4}_{2.4}$ &0.44/21\\
LMC field north&8.69&Bc&50&(-13.7,-11.0) &34&6& 2.2$\pm^{3.2}_{2.6}$ &0.43/21\\
LMC field north&8.69&Bd&50&(-13.7,-11.0) &34&6& 2.8$\pm^{2.6}_{3.2}$ &0.43/21\\
LMC field north&8.69&Be&50&(-13.7,-11.0) &34&6& 2.0$\pm^{3.4}_{2.4}$ &0.43/20\\
LMC field north&8.69&Bf&50&(-13.7,-11.0) &34&6& 2.6$\pm^{2.8}_{3.0}$ &0.42/21\\
Supergiant Shell LMC\,4 north&1.81&Ha&30&(-13.8,-11.0)&35&4&1.6$\pm^{3.0}_{1.8}$&0.86/10\\
Supergiant Shell LMC\,4 north&1.81&Hb&30&(-13.8,-11.0)&35&4&2.2$\pm^{2.4}_{2.4}$&0.82/10\\
Supergiant Shell LMC\,4 north&1.81&Hc&30&(-13.8,-11.0)&35&4&1.4$\pm^{3.2}_{1.6}$&0.85/10\\
Supergiant Shell LMC\,4 north&1.81&Hd&30&(-13.8,-11.0)&35&4&1.6$\pm^{3.0}_{1.8}$&0.82/10\\
Supergiant Shell LMC\,4 north&1.81&He&30&(-13.8,-11.0)&35&4&1.0$\pm^{3.6}_{1.2}$&0.89/10\\
Supergiant Shell LMC\,4 north&1.81&Hf&30&(-13.8,-11.0)&35&4&4.0$\pm^{0.6}_{4.2}$&0.93/10\\
Supergiant Shell LMC\,4 north&1.81&Ba&30&(-13.8,-11.0)&35&4&3.0$\pm^{1.6}_{3.2}$&0.37/10\\
Supergiant Shell LMC\,4 north&1.81&Bb&30&(-13.8,-11.0)&35&4&3.8$\pm^{0.8}_{4.0}$&0.48/10\\
Supergiant Shell LMC\,4 north&1.81&Bb&30&(-13.8,-11.0)&35&4&3.8$\pm^{0.8}_{4.0}$&0.48/10\\
Supergiant Shell LMC\,4 north&1.81&Bc&30&(-13.8,-11.0)&35&4&2.8$\pm^{1.8}_{3.0}$&0.36/10\\
Supergiant Shell LMC\,4 north&1.81&Bd&30&(-13.8,-11.0)&35&4&3.8$\pm^{1.0}_{2.8}$&0.51/10\\
Supergiant Shell LMC\,4 north&1.81&Be&30&(-13.8,-11.0)&35&4&2.6$\pm^{2.0}_{2.6}$&0.44/10\\
Supergiant Shell LMC\,4 north&1.81&Bf&30&(-13.8,-11.0)&35&4&3.4$\pm^{1.2}_{3.6}$&0.42/10\\
LMC field south&6.44&Ha&50&(-13.2,-11.0)&17&11&1.8$\pm^{2.6}_{0.6}$&0.89/10\\
LMC field south&6.44&Hb&50&(-13.2,-11.0)&17&11&3.2$\pm^{2.2}_{1.6}$&0.81/10\\
LMC field south&6.44&Hc&50&(-13.2,-11.0)&17&11&1.6$\pm^{2.4}_{0.6}$&0.89/10\\
LMC field south&6.44&Hd&50&(-13.2,-11.0)&17&11&2.2$\pm^{3.0}_{0.8}$&0.86/10\\
LMC field south&6.44&He&50&(-13.2,-11.0)&17&11&2.2$\pm^{1.6}_{1.2}$&0.94/10\\
LMC field south&6.44&Hf&50&(-13.2,-11.0)&17&11&2.0$\pm^{3.0}_{0.6}$&0.85/10\\
LMC field south&6.44&Ba&50&(-13.2,-11.0)&17&11&2.2$\pm^{1.4}_{0.8}$&1.19/10\\
LMC field south&6.44&Bb&50&(-13.2,-11.0)&17&11&3.2$\pm^{1.6}_{1.6}$&1.20/10\\
LMC field south&6.44&Bc&50&(-13.2,-11.0)&17&11&2.0$\pm^{1.4}_{0.8}$&1.21/10\\
LMC field south&6.44&Bd&50&(-13.2,-11.0)&17&11&3.0$\pm^{1.6}_{1.2}$&1.24/10\\
LMC field south&6.44&Be&50&(-13.2,-11.0)&17&11&2.0$\pm^{1.2}_{0.6}$&1.31/10\\
LMC field south&6.44&Bf&50&(-13.2,-11.0)&17&11&2.6$\pm^{1.6}_{1.0}$&1.24/10\\
     \noalign{\smallskip}
     \hline
     \end{tabular}
     \end{flushleft}
Note: (a) number of all selected AGN; 
      (b) number of AGN with LMC \HI\ columns $>10^{21}\ {\rm cm^{-2}}$
     \label{tab:gasfrac}
\end{table*}

For a consistency check with the result derived from X-ray spectral 
fitting of individual AGN (cf. Paper\,I) we compared the $N_{\rm H}$ model 
we have derived from the $\log N\ -\ \log S$ analysis with the $N_{\rm H}$ 
values we have derived from X-ray spectral fitting. There is within the 
uncertainties consistency for both $N_{\rm H}$ models, which gives this 
result credibility.

To investigate regional variations to the amount of gas additional to the 
\HI\ we considered a few special areas inside the total LMC field. These are 
North, South, and LMC\,4 (see Fig.~\ref{ps:imalmcreg}). In each we constructed
the $\log N\ -\ \log S$ and investigated whether it deviated from the 
overall $\log N\ -\ \log S$.

\begin{figure}
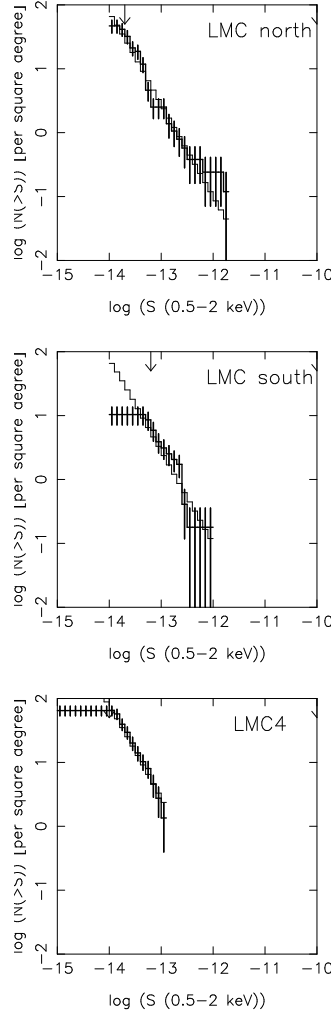
 
  \centering{
  \vbox{\psfig{figure=MS1834f6a.eps,width=4.2cm,angle=-90.0,%
  bbllx=3.5cm,bblly=4.0cm,bburx=19.5cm,bbury=18.6cm,clip=}}\par
  \vbox{\psfig{figure=MS1834f6b.eps,width=4.2cm,angle=-90.0,%
  bbllx=3.5cm,bblly=4.0cm,bburx=19.5cm,bbury=18.6cm,clip=}}\par
  \vbox{\psfig{figure=MS1834f6c.eps,width=4.2cm,angle=-90.0,%
  bbllx=3.5cm,bblly=4.0cm,bburx=19.5cm,bbury=18.6cm,clip=}}\par
            }
  \caption[]{$\log N\ -\ \log S$ of the observed AGN sample in the 
  field of the northern LMC (upper panel), southern LMC (middle panel), 
  and Supergiant Shell LMC\,4 (lower panel) for the spectrally hard 
  band (model Hd, cf. Tab.\,3) which has been observed with off-axis angles 
  $<$19.\arcmin5. The $\log N\ -\ \log S$ has been corrected for 
  incompleteness due to the variable exposure across the LMC field, the 
  absorption due to galactic and LMC gas (the \HI), and  
  the incomplete sampling of AGN with a given number of counts in the source 
  circle. The chosen threshold was 50 counts for the sample of the 
  northern and southern LMC and 30 counts for the AGN sample in the 
  LMC\,4 area. The best-fit has been found for $N_{\rm H}$ 
  scale factors of 4.2, 3.2, and 2.6 respectively for the same high column 
  regime of the LMC gas. The thin solid histogram gives the theoretical 
  $\log N\ -\ \log S$ (model B of GSH01) and the arrow marks the lower flux 
  threshold used in the least-square fit.}
  \label{ps:lognlmcns4}
\end{figure}

The field of the northern LMC is less affected by hot diffuse gas than the 
southern field of the LMC which contains the 30\,Dor complex (cf. Fig.\,4 and 
Fig.\,1 in Paper\,I). This region is therefore better suited to derive the 
$\log N\ -\ \log S$ in the LMC area of the sky. It contains 34 AGN with more 
than 50 detected counts. This field contains the Supergiant Shell LMC\,4 for 
which the $\log N\ -\ \log S$ will also be derived in this section. Additional
$N_{\rm H}$ to the \HI\ in the high column ($>10^{21}\ {\rm cm}^{-2}$) regime
of the LMC gas with a factor of $2.8\pm^{2.6}_{2.5}$\,$N_{\rm H}$ for
the spectrally hard band (cf. Fig.~\ref{ps:lognlmcns4}, upper panel) and of 
$2.3\pm^{3.1}_{2.7}$\,$N_{\rm H}$ for the spectrally broad band is required 
for the $\log N\ -\ \log S$ in the northern LMC. Although this factor is 
uncertain this result may mean that additional $N_{\rm H}$ with a similar 
amount as determined from the overall \HI\ map may exist in the field of the 
northern LMC.

The field of the southern LMC contains the 30\,Dor complex with copious 
diffuse X-ray emission and the western complex of large \HI\ column densities 
(cf. Fig.\,4 and Fig.\,1 of Paper\,I). Further south of the large cloud 
complex the  \HI\ columns decrease dramatically and background X-ray sources 
become visible. We constructed the $\log N\ -\ \log S$ in this part of the 
LMC. We excluded the complex of diffuse X-ray emission (Diffuse~4) for reasons
discussed in Sect.\,3.3. We detected 17 AGN in this field with more than 50 
observed counts. Additional gas to the \HI\ by an amount of 
$2.2\pm^{2.5}_{1.0}$\,$N_{\rm H}$ for the spectrally hard band 
(cf. Fig.~\ref{ps:lognlmcns4}, middle panel) and of 
$2.5\pm^{1.5}_{1.1}$\,$N_{\rm H}$ for the spectrally broad band is required 
for the $\log N\ -\ \log S$ in the southern LMC. Although this factor is 
uncertain, this result means that additional $N_{\rm H}$ likely exists in the 
field of the southern LMC. This additional gas is within the uncertainties 
comparable to the additional gas found for the northern LMC.

The northern field of the Supergiant Shell LMC\,4 has been observed with 
a large integrated exposure of at least a few 10~ksec and up to $\sim$80~ksec.
In Paper\,II a catalog of X-ray sources in this field of the LMC has been
generated applying a maximum likelihood detection procedure to the merged 
data of this field. 35 of the sources detected in this field and observed 
with more than 30 counts were classified as background sources. We 
corrected
for the incompleteness of the chosen AGN sample which amounts to 70\%.
These sources were used for a $\log N\ -\ \log S$ analysis. We generated the 
observed $\log N\ -\ \log S$ distribution which we corrected for the 
variable exposure and gas absorption and for the incomplete sampling of 
AGN with 30 counts in the source circle. From the $\log N\ -\ \log S$ 
analysis for this observation we derived gas additional to the LMC \HI\ by 
a factor of $2.0\pm^{2.8}_{2.4}$ at 90\% confidence in the spectrally hard 
band and for the high column ($>10^{21}\ {\rm cm}^{-2}$) regime of the LMC 
gas (cf. Fig.~\ref{ps:lognlmcns4}, lower panel). This value is uncertain 
but consistent with some gas additional to the \HI\ in the high column 
regime of the LMC gas. Four of the candidate AGN are located in regions
of high LMC columns. The area of this high column regime is associated 
with the western high column boundary of the Supergiant Shell where CO 
emission has also been detected (Yamaguchi et al. 2001). This could be an 
indication for the existence of molecular hydrogen.

We have derived constraints for the LMC gas from the background AGN in 
fields of the LMC with a size of a few square degrees. 

In the fields for which we derived constraints a wide range of LMC column 
densities are involved, likely variable molecular mass fractions, and 
different contributions of warm and hot diffuse gas. Thus only mean properties
of the LMC gas could be derived. In addition the analysis is complicated by 
hot diffuse gas (mainly reducing the effective area where AGN can be 
detected). In addition, for the investigated fields, it cannot be guaranteed 
that the selected AGN sample is complete, i.e. that all background X-ray 
sources with an intrinsic number of counts above the chosen counts threshold 
have been considered. Especially in regions with a high integrated exposure 
(e.g. in the field of the Supergiant Shell LMC4) faint AGN are easily observed
with a few 10 counts. For the X-ray sources detected in the LMC field we also 
followed the classification scheme given in Haberl \& Pietsch (1999) and in 
most cases we did not consider X-ray sources which have a close foreground 
star.

Thus we have performed a simulation of the observed AGN in the LMC field. We 
used the $N_{\rm H}$ model given by the {\sl Parkes} \HI\ map. From such an 
analysis it follows that the simulated number of AGN and clusters in the LMC 
field is $\sim$30\% larger than the observed number of AGN. This would mean 
that 24 AGN and clusters of galaxies have not been found in our analysis. This 
would be a considerable number. In a second step we used in the simulation for 
the $N_{\rm H}$ model the {\sl Parkes} \HI\ map which we scaled in the high 
column ($> 10^{21}\ {\rm cm}^{-2}$) regime of the LMC gas by a factor of two. 
From this simulation we derived a number of AGN which is in agreement with 
the number of AGN observed in the investigated fields of the LMC (cf. 
Fig.~\ref{ps:figsim1} for a comparison of the simulated with the observed 
AGN sample). This result indicates that absorbing gas additional to the LMC 
\HI\ by about a factor of two could be present in the high column regime 
of the LMC gas.

\subsection{Dependence on the cluster of galaxy component}

In Sect.\,2 we have introduced a two component model for the 
$\log N\ -\ \log S$ comprising an AGN and a cluster component. We reproduced 
the AGN and the cluster component of the $\log N\ -\ \log S$ from Fig.\,3 of 
GRS99 and Fig.\,3 of GSH01. In the analysis of Sect.\,4.1 we have assumed that
this description of the $\log N\ -\ \log S$ is also valid for the LMC field. 
But clusters of galaxies and galaxy groups may show variations across the sky 
(cf. Giuricin et al. 2000). We thus will, in this section, explore the effect 
of a varying cluster component in the $\log N\ -\ \log S$ on the derived 
amount of intervening LMC gas. We therefore introduce a cluster scaling 
factor which we will determine in a least-square fit of the observed 
$\log N\ -\ \log S$ with the theoretical $\log N\ -\ \log S$.

There may be one point of concern with respect to the cluster component.
We originally selected the AGN sample excluding objects with a large
extent likelihood ratio ($ML_{\rm ext}>30$). Our selection does not 
contain clusters with appreciable extension. This fact is supported by 
the fact that we found in our sample no strongly extended source. But 
clusters have been found to be extended in the \ros\ sample. Rosati et al. 
(1995) e.g. have investigated a $\sim$3 square degree field of deep \ros\ 
{\sl PSPC} observations. They found $\sim$13 clusters in their sample 
which have a significant extension in excess of the point-spread function 
of the {\sl PSPC}. About 10 of these clusters have an extension of 
$\sim$1\arcmin. Such clusters may not have been included in our AGN sample 
as we selected against objects with a significant extension.

Our sample closely matches the sample of point-like X-ray sources for which 
the HBG98 $\log N\ -\ \log S$ is valid. Note that also Rosati et al. (1995) 
set up a sample of point-like X-ray sources in their sample. They found that 
this sample is in agreement with the HBG98 $\log N\ -\ \log S$. 
We thus suspect that only weaker and unresolved clusters contribute to our 
selected AGN sample. For a more complete cluster sample we would have to 
include the significantly extended X-ray sources as well. But such a sample 
could be confused with diffuse LMC structure and therefore may be difficult 
to confine in areas where diffuse LMC X-ray emission exists.

Therefore a fit of the ``flattened'' $\log N\ -\ \log S$ (for model Hd) with 
an AGN and a cluster of galaxy component (keeping the AGN scaling factor 
fixed to 1.0) has been performed to determine the cluster scaling factor as
well as the amount of gas additional to the \HI\ . For the investigated sample
we derived a cluster scaling factor of $1.0\pm^{1.5}_{1.0}$ at 90\% confidence 
which is consistent with a cluster scaling factor of 1.0. The amount of gas 
additional to the \HI\ is derived to be $2.2\pm^{3.2}_{2.6}$ but remains 
unconstrained due to the large uncertainties. In a next step a fit has
been performed in which the $N_{\rm H}$ scale factor and the AGN scale factor
have been set to a value of 1.0. For the cluster scale factor which has been 
varied in the fit a value in the range $0.0 - 0.9$ at 90\% confidence has 
been found. 

The interpretation of this result could mean that there is no large amount of 
additional gas in the high column regime of the LMC gas but that a significant
fraction of the clusters of galaxies which are according to the cluster 
$\log N\ -\ \log S$ expected to exist in the field of the LMC have not been 
found. To further explore this possibility we performed a simulation with a 
theoretical $\log N\ -\ \log S$ where we set the cluster scaling factor to 
0.0, 0.2, and 0.5 respectively. We assumed absorption by LMC gas as reproduced
in the {\sl Parkes} \HI\ map. From this simulation we derived a number of 40, 
52, and 64 AGN and clusters respectively to exist in the observed LMC field 
excluding the region around 30~Dor. In case of a cluster scaling factor of 
0.2 this number would be consistent with the number of 50 observed AGN in the 
same field. In addition the distribution of LMC absorbing column densities 
and source counts derived for the simulated AGN and cluster sample for this 
case matches closely the observed distributions (cf. Fig.\,2, {panel~d). 
Thus 
there appears to exist the possibility that a cluster of galaxy component 
which is reduced by about a factor of five can account for the deficiency in 
the observed $\log N\ -\ \log S$. If this interpretation is correct 
then there remains the question whether the fraction of clusters of galaxies 
is reduced in the field of the LMC by a considerable fraction or whether a 
large fraction of clusters of galaxies in the field of the LMC has not 
been included in our sample of background sources (cf. the catalog of 
background sources used in our analysis as given in Paper\,II).

\section{Nature of the absorbing gas additional to the HI}

In the previous section we have derived constraints on the amount of
absorbing gas in the field of the LMC from the $\log N\ -\ \log S$ analysis 
of background sources. We could not determine whether such a gas is in the
atomic, molecular, or dusty phase.

\subsection{Molecular gas}

Here we will discuss the possibility that such additional gas is molecular, we
will constrain the molecular mass fraction of such a gas, and we will compare 
our result with estimates of the amount of molecular gas in the LMC inferred 
from other information. The amount of molecular hydrogen $N_{\rm H_{2}}$ can 
be estimated from the total amount of absorbing gas $N_{\rm H}^{\rm tot}$ and 
the amount of atomic hydrogen $N_{\rm \sHI}$ as

\begin{equation}
  N_{\rm H_{2}} = \frac{1}{2.2}\big(N_{\rm H}^{\rm tot} 
  - N_{\rm \sHI\ }\big)
\end{equation}

\noindent
assuming that in the \ros\ {\sl PSPC} band the photoabsorption cross section 
of molecular hydrogen corrected for the helium contribution is a factor of 
2.2 larger than that of atomic hydrogen (cf. Appendix~A). The molecular mass 
fraction $f$ can be determined from the equation

\begin{equation}
  f = \frac{N_{\rm H}^{\rm tot} - N_{\rm \sHI\ }}{N_{\rm H}^{\rm tot} + 0.1\ N_{\rm \sHI\ }}
\end{equation}

\noindent
(see also Eq.\,1 and Eq.\,2 in Paper\,I which have been derived taking only 
the contribution of hydrogen to the photoabsorption cross section into 
account).

We have derived in the previous section from the spectrally hard band that 
$N_{\rm H}^{tot} = 2.9\pm^{3.3}_{1.6}\ N_{\rm \sHI}$ assuming that the
metallicity of the ISM of the LMC is $-$0.3~dex lower than the metallicity
of the galactic ISM. This implies that the mean molecular mass fraction is 
constrained to $f = 0.63\pm^{0.20}_{0.42}$ assuming that the gas additional 
to the \HI\ is molecular.

We can compare this result with the molecular mass fraction of the gas in 
the Magellanic Clouds derived by Richter (2000) from UV absorption 
measurements in the direction of 7 stars in  the LMC and the SMC. He found 
a low value (less than 10\%) for the molecular mass fraction which is 
depending on the value of the hydrogen column density. Only in regions of 
hydrogen columns $> 10^{21}\ {\rm cm}^{-2}$ a molecular mass fraction 
$\approxgt$1\% has been derived. From a recent {\sl FUSE} survey towards
interstellar sight lines in the LMC a small mean value for the diffuse
molecular hydrogen of $\sim$2\% has been derived (Tumlinson et al. 2001). 
But from the same analysis a higher value for the diffuse molecular mass 
fraction of $\sim$10\% is found in the high column ($>10^{21}\ {\rm cm}^{-2}$)
regime of the LMC gas. Savage et al. (1977) derived from measurements with 
the Copernicus satellite for a large sample of galactic stars the molecular 
mass fraction. For comparison we made use of the dependence of $f$ on 
$N_{\rm H}$ which has been found in their analysis. We determined the mean 
molecular mass fraction $<f>$ weighted with the $N_{\rm H}$ distribution of 
our candidate AGN sample (but we did not scale this $N_{\rm H}$ with the scale
factor we have derived). We found a value of $<f> = 0.12$. If we restricted 
the analysis to AGN observed in the high column 
($N_{\rm H}^{\rm LMC} \ge 10^{21}\ {\rm cm}^{-2}$) regime then we derive for 
the molecular mass fraction a mean value of $<f> = 0.20$. This value would be 
within the uncertainties lower than the molecular mass fraction we have 
derived for the high column ($>10^{21}\ {\rm cm}^{-2}$) regime of the LMC gas 
making use of the $\log N\ -\ \log S$ of our AGN sample.

Still we cannot exclude that the additional gas is in part warm diffuse 
gas. But if we compare with the Andromeda galaxy (M31) with an \HI\ mass 
of $4\ 10^9\ M_{\odot}$ (Braun 1991) and a maximum total mass for the 
hot diffuse gas of $(1.0\pm0.3)\ 10^6\ M_{\odot}$ (Supper et al. 2001) then 
a similar contribution of hot diffuse gas as in M31 would give for the LMC 
a maximum mass for the hot diffuse gas of $(1.0-1.5)\ 10^5\ M_{\odot}$.
But the LMC is in comparison to M31 a star forming galaxy with a higher
fraction of hot diffuse gas (about a factor of 10).
Thus a similar amount of hot diffuse gas as in M31 may be present in
the LMC. Such a mass is still small compared to the mass of the neutral 
gas in the LMC disk of $5.2\ 10^8\ M_{\odot}$ (Kim et al. 1998).

Thus it is likely that such additional gas would be molecular. 
The {\sl Columbia} {\sl Southern} {\sl Telescope} has been used to 
perform a 6\D$\times$6\D 
survey of the LMC field (Cohen et al. 1988). Adopting a conversion
factor $X_{\rm LMC} = N_{\rm H_{2}} / I_{\rm CO}$ between the molecular 
column density $N_{\rm H_{2}}$ and the absolute CO intensity $I_{\rm CO}$ 
for the LMC gas, which is a factor of 6 larger than the galactic value 
$X_{\rm G}$ (which follows from the finding that the LMC CO complexes appear 
to be underluminous in CO compared to the galactic molecular cloud complexes), 
a total molecular mass has been derived in the high column 
($> 10^{21}\ {\rm cm}^{-2}$) regime of the LMC gas, which is 30\% of the 
neutral hydrogen mass (cf. Rubio 1999). 

From the $^{12}$CO {\sl NANTEN} LMC survey of the LMC a molecular mass 
fraction of (10--20)\% has been derived in the high column 
($> 10^{21}\ {\rm cm}^{-2}$) regime of the LMC gas assuming a somewhat 
smaller $X_{\rm LMC}$ factor (2--4 times the galactic value), which has 
been derived under the assumption that the CO clouds in the LMC are
virialized (Mizuno et al. 2001).
 
Israel (1997) derived from far-infrared and \HI\ data a global mean molecular 
mass fraction of $\sim$0.2 for the LMC gas. In addition, for individual CO 
clouds with \HI\ column densities in the range $(0.7-3.5)\ 10^{21}\ 
{\rm cm^{-2}}$ in the mean, a larger molecular mass fraction of $\sim$0.4 
is found.

\subsection{Obscuration by absorbing clouds}

We test the hypothesis that at least part of the deficiency in the 
$\log N\ -\ \log S$ of background sources in the LMC field is due 
to obscuration by dark clouds. A fractal size distribution of the \HI\ clouds 
has recently been found by Stanimirovic et al. (1999) from a {\sl Parkes} and
{\sl ATCA} survey of the SMC. In addition Stanimirovic (2000) has found that 
the size distribution of dust column density fluctuations in the SMC is 
described by the same relation as for the \HI\ clouds.

We now assume that there exists a powerlaw size distribution of dark 
obscuring clouds in the LMC with a powerlaw distribution in the size. The 
projected fractal dimension of $D_{\rm p}=1.5$ (Stanimirovic et al., 1999)
is similar to values found for molecular clouds in the Milky Way in the size 
range $\sim$0.05 to 100 pc (e.g. Falgarone, Phillips \& Walker 1991). And also
for the LMC Kim et al. (1999) have found that the size distribution of \HI\ 
shells is consistent with a powerlaw distribution with a slope of 
$-1.5\pm0.4$. The distribution of cloud sizes for a projected fractal 
dimension of $D_{\rm p}=1.5$ follows then a powerlaw with a slope of $-2.5$ 
(e.g. Stanimirovic et al., 1999).

\begin{equation}
  \frac{dN}{dL} = \ C\ L^{-(D_{\rm p} + 1)}
\end{equation}

\noindent
with $dN$ the number of clouds per cloud size interval $dL$, $L$ the 
size of the cloud, $D_{\rm p} = 1.5$, and the scaling constant $C$.
For the number of clouds with a size $\lambda > L$ one derives

\begin{equation}
  N (\lambda > L) = \frac{C}{D_{\rm p}} L^{-D_{\rm p}}
\end{equation}

To apply this relation for the LMC we simply inspect the \ros\ {\sl PSPC} 
image of a 10\D $\times$ 10\D\ field of the LMC in the energy range 
(0.4 - 1.3 keV, cf. Fig.~1 in Paper\,I). We clearly can recognize three large 
dark roughly elliptical clouds which have areas in a narrow range around 
$A_{\rm large}\sim$0.45 square degrees and a mean extent (in degrees) of 
$<L_{\rm large}> = \sqrt {\frac{4 A_{\rm large}}{\pi }} \sim 0.76.$ For a 
distance to the LMC of 50\,kpc we derive for these large dark clouds a mean 
cloud extent of 660\,pc. We then can determine the scaling constant $C$ from

\begin{equation}
  C = D_{\rm p}\ \frac{N_{\rm large}}{\big(L_{\rm large}\big)^{-D_{\rm p}}}
\end{equation}

\noindent
with $N_{\rm large} = 3$, $L_{\rm large}$ 
= 0.76\D\ and $D_{\rm p} = 1.5$ 
we obtain

\begin{equation}
  N(\lambda > L)\ =\ 2.98\ L^{-1.5}
\end{equation}

We now determine the area covered by the powerlaw distribution of dark
clouds in the LMC field, for which we determined the large size end. 
We use for the area of a single cloud of size $L$ the expression 
$A_{\rm cloud}\ =\ \frac{\pi}{4}\ L^2$ and determine the total area $A$ 
of all dark clouds as

\begin{equation}
  A\ = \int A_{\rm cloud}\ {\frac{dN}{dL}\ dL}\ 
  = \frac{\pi}{4}\ C  \int^{L_2}_{L_1} L^{-D_{\rm p} + 1} dL 
\end{equation}

Solving the integral and using the expression for $C$ from Eq.\,20 
then we obtain

\begin{equation}
  A = \frac{\pi }{4}\
      \frac{D_{\rm p}}{(2-D_{\rm p})}\ 
      \frac{N_{\rm large}}{\big(L_{\rm large}\big)^{-D_{\rm p}}}\ 
      \big( L_2^{2-D_{\rm p}} - L_1^{2-D_{\rm p}}\big)
\end{equation}

If we use the projected fractal dimension $D_{\rm p} = 1.5$ and if we use 
a lower cutoff for the cloud size in this distribution with a value of 
$L_{1}$=30\,pc (equivalent to a size of 2\arcmin) then we find the 
total area covered by dark clouds to 2.5 square degrees. Assuming this 
fractal cloud distribution extends over a 5\D\ radius of the LMC then we 
can derive the fractional area which is obscured by these dark clouds. We 
find a fraction of only 3\%. We have made the assumption that there 
exists a cutoff in the size of dark clouds with a lower value of 30\,pc. As 
can be seen from Eq.\,23 this value will not change much if we decrease 
the lower size limit. The asymptotic limit for setting the lower cloud size
to $0$ is 3.1 square degrees. We will argue below that in any case a 
lower limit for the size of dark obscurating clouds of less than $\sim$30\,pc 
appears to be unphysical.

If we put forward the interpretation for the deficit of AGN in 
the $\log N\ -\ \log S$, obscuration by a powerlaw distribution
of dark obscurating clouds, then we apparently do not succeed to explain 
a deficit of at least a few 10\% as apparently is required from the 
$\log N\ -\ \log S$. There is still the possibility that we underestimate
the number of dark clouds with sizes larger than 660\,pc in the LMC field with
a radius of 5 degrees. But the number of such large clouds would have 
to be at least 5 times larger (i.e. $\sim$15) to be able to account for the 
deficit of AGN in the $\log N\ -\ \log S$.

We so consider it as unlikely that a powerlaw distribution of dark clouds
with a projected fractal dimension $D_{\rm p} = 1.5$ alone is responsible 
for the forementioned deficit.

How realistic is it, from a physical point of view, that dark cloud 
obscuration plays a role? Assuming densities as are observed in molecular 
clouds of $\sim$$10^2\ {\rm cm}^{-3}$ then for a spherical cloud with radius 
$R = 30 {\rm pc}$ and a mean crossing length of 1.27\,$R$ we derive a mean 
column density of $1.2\ 10^{22}\ {\rm cm}^{-2}$. Such a cloud is opaque to 
soft X-rays. A somewhat smaller cloud of size 5\,pc has a mean column density 
of $2.0\ 10^{21}\ {\rm cm}^{-2}$ and would be transparent above 0.5~keV,
i.e. to soft X-rays. This means a cutoff size smaller than 30\,pc for dark 
obscurating clouds would not be physical for such densities found in molecular 
clouds.

\section{Summary and conclusions}

We constructed the $\log N\ -\ \log S$ of background X-ray sources in 
the field of the LMC (excluding the region of extended diffuse X-ray emission 
in the 30\,Dor complex) observed with the \ros\ {\sl PSPC} and published in 
the catalog of HP99. We only considered X-ray sources which were observed in 
the inner 20\arcmin\ of the {\sl PSPC}, which had at least 50 observed counts
and which have been classified in Paper\,II as background X-ray sources.
We corrected this observed $\log N\ -\ \log S$ for incompleteness due to 
the variable exposure across the LMC field and the varying absorption due to 
the LMC gas.

In a first step we compared the observation derived $\log N\ -\ \log S$ 
with the theoretical $\log N\ -\ \log S$ of the SXRB as given in GRS99 
which comprises, besides an AGN, a cluster component. From this comparison 
it is found that the observed $\log N\ -\ \log S$ has a deficiency with 
respect to the $\log N\ -\ \log S$ of the SXRB. We investigated several 
possibilities to explain this deficiency.

One explanation for this deficiency could be absorption of the background 
sources by gas additional to the measured \HI\ in the high column ($>10^{21}\ 
{\rm cm}^{-2}$) regime of the LMC gas. We would derive gas additional to the 
\HI\ in the high column ($>10^{21}\ {\rm cm}^{-2}$) regime in the field of 
the LMC by a factor of $1.9\pm^{3.3}_{1.6}$ at 90\% confidence assuming that 
the metallicity of the ISM of the LMC is $-$0.3~dex lower than the metallicity 
of the galactic ISM. If this additional gas is molecular then a molecular mass
fraction of $f = 0.63\pm^{0.20}_{0.42}$ would be derived assuming that the 
``effective'' photoabsorption cross section of molecular hydrogen is a factor 
of 2.2 larger than that of atomic hydrogen (taking the cross section of helium
into account but neglecting the contribution of other elements and molecules).
We also applied this analysis to the AGN observed in the northern LMC, the 
southern LMC, and the northern part of the Supergiant Shell LMC~4 and we 
derived for these regions an amount of gas additional to the \HI\ which is 
consistent with the amount of gas derived for the whole LMC field.

An alternative interpretation for the source deficiency would be that the 
$\log N\ -\ \log S$ in the field of the LMC deviates from the theoretical 
$\log N\ -\ \log S$ which we have used in the investigation. Especially the 
$\log N\ -\ \log S$ of the clusters of galaxies could be different in the 
field of the LMC. In order to account for the observed $\log N\ -\ \log S$
the $\log N\ -\ \log S$ of galaxy clusters would have to be largely reduced 
by a factor of $\sim$5. Another explanation could be that our selected sample 
of clusters of galaxies is not complete.

The result of this analysis is that a $\log N\ -\ \log S$ analysis of 
background X-ray sources in the field of the LMC can, in principle, be used 
to constrain gas additional to the measured \HI\ in such a galaxy. A source 
classification scheme has to be applied. It also is required that the 
completeness of the selected AGN sample and the clusters of galaxies sample
is guaranteed, which is difficult to achieve in regions with extended diffuse 
X-ray emission. Also it has to be assumed that the theoretical
$\log N\ -\ \log S$ applies to the field of the LMC. If all of these 
assumptions are fulfilled then from the amount of additional gas constraints 
on the amount of molecular gas can be inferred in case all additional gas is 
molecular (assuming that the amount of hot ($>10^6$~K) diffuse gas and of 
cold dust compared to molecular gas is small).

\appendix

\section{The photoabsorption cross section of molecular hydrogen}

The results derived in the present paper are in part based on the assumptions
made on the photoabsorption cross section of molecular hydrogen. Recently,
Yan, Sadeghpour, \& Dalgarno (1998) have derived an analytical description 
of the photoabsorption cross section of molecular hydrogen which extends 
from 15.4~eV to above 85~eV (into the keV regime). In comparing with the
photoabsorption cross section of atomic hydrogen they concluded that the
cross section of molecular hydrogen is about a factor of 2.8 larger (which
means that the cross section per H-atom is 1.4 larger for molecular hydrogen).
WAM used the description for the photoabsorption cross section of Yan, 
Sadeghpour, \& Dalgarno (1998) but they applied a modification at energies 
below 85~eV. They also found that the cross sections of molecular and atomic 
hydrogen differ by about a factor of 2.85. 

The differences of the energy dependence of the molecular and atomic cross 
sections are not very pronounced and in most cases the quality of the 
measured spectra is not that high that large differences in the columns 
derived from X-ray spectral fitting are obtained.

Hydrogen is the most abundant element assuming cosmic abundances.
Helium is the second abundant element and has to be taken into account
if absorption by gas in the energy range of soft X-rays is considered.
All other elements have much lower abundances and their contribution
accordingly is of less importance. In order to estimate the effect of
molecular hydrogen on the effective photoabsorption cross section we
take in addition helium into account.

\begin{figure} 
  \centering{
  \vbox{\psfig{figure=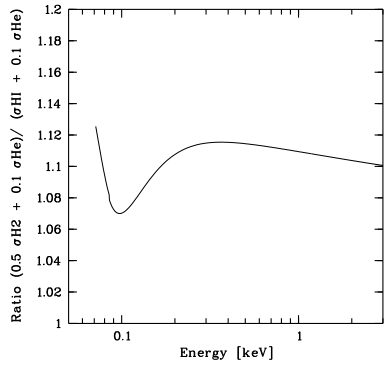,width=5.5cm,angle=0.0,%
  bbllx=3.5cm,bblly=1.5cm,bburx=16.0cm,bbury=13.0cm,clip=}}\par
            }
  \caption[]{Ratio of photoabsorption cross sections 
  $(0.5 \sigma H_{2} + 0.1 \sigma He) / (\sigma \HI\ + 0.1 \sigma He)$
  of molecular ($H_{\rm 2}$) and atomic (\HI ) hydrogen. The description 
  for the photoabsorption cross sections of WAM has been used.}
  \label{ps:sighihe}
\end{figure}

We made use of the description of the cross section of molecular hydrogen
and of helium as given in WAM, see also  Yan, 
Sadeghpour, \& Dalgarno (1998) and we assume for helium an abundance of 
10\% of hydrogen. We determined the ratio of the effective cross sections 
$(0.5 \sigma H_{2} + 0.1 \sigma He) / (\sigma \HI\ + 0.1 \sigma He)$ which
we show in Fig.~\ref{ps:sighihe}. This ratio is $\sim$1.10 for energies 
above 0.1~keV. This result means that the effective cross section of 
molecular hydrogen is about 2.2 times the cross section of atomic hydrogen. 
If we neglect the contribution of metals to the photoabsorption cross section
(this assumption may be valid for the LMC where the metal abundance is by a 
factor of two lower than in the Galaxy) then we can derive the column of 
molecular hydrogen from Equ.\,16.

\acknowledgements
The \ros\ project is supported by the Max-Planck-Gesellschaft and the 
Bundesministerium f\"ur Forschung und Technologie (BMFT). We thank J. Kerp
for reading an earlier version of the manuscript and an anonymous referee for
helpful comments. PK is supported by the Graduiertenkolleg on the 
``Magellanic Clouds and other Dwarf galaxies'' (DFG GRK\,118).

\end{document}